\documentclass[10pt]{article}

\usepackage[
paperwidth=493.228pt,
paperheight=700.157pt,
left=45pt,
right=47.727pt,
top=42pt,
bottom=45pt
]{geometry}
\usepackage{mathtools,amsmath,amssymb,dsfont,bm}
\usepackage{graphicx}
\usepackage{natbib}
\usepackage[protrusion=true,expansion=false]{microtype}
\usepackage{hyperref}
\usepackage{enumitem}
\usepackage{caption}

\hypersetup{
    colorlinks=true,
    linkcolor=black,
    citecolor=black,
    urlcolor=black
}

\def\bs{\ensuremath\boldsymbol}

\def\eps{\ensuremath\mathrm{\varepsilon}}

\def\rhat{\ensuremath\bs{\hat{r}}}
\def\that{\ensuremath\bs{\hat{\theta}}}

\def\psitil{\ensuremath\tilde{\psi}}
\def\zetatil{\ensuremath\tilde{\zeta}}

\def\psihat{\ensuremath\hat{\psi}}
\def\zetahat{\ensuremath\hat{\zeta}}

\def\zhat{\ensuremath\bs{\hat{z}}}

\def\onab{\ensuremath\overline{\nabla}}

\newcommand{\BVfreq}{{Brunt--V\"ais\"al\"a frequency}}

\title{\textbf{A multiscale theory of convective momentum transport in the tropical atmosphere}}
\author{%
Edward J. Goldsmith$^{1}$, Levi Dittman$^{1}$, and Joseph A. Biello$^{2}$\\[0.6em]
\small $^{1}$Department of Mathematics, Hillsdale College, Hillsdale, MI, USA\\
\small $^{2}$Department of Mathematics, University of California, Davis, Davis, CA, USA\\[0.6em]
\small Corresponding author: \texttt{egoldsmith@hillsdale.edu}
}
\date{}

\begin{document}
\maketitle
\begin{abstract}
	A multiple-scales asymptotic framework is employed to analyze the interaction between a field of small-scale convective circulations and the large-scale tropical circulation. Closed equations governing the large-scale flow are derived, providing a systematic closure for the influence of unresolved convection. The influence of unresolved convection enters through a non-local momentum diffusion operator that provides a first-principles representation of convective momentum transport and naturally couples the vertical baroclinic modes of the large-scale circulation. The classical Matsuno--Gill model is recovered as a limiting approximation in which the derived momentum transport operator reduces to the phenomenological damping introduced by Gill, thereby placing the classical theory on a systematic first-principles foundation. Two applications of the generalized theory are considered. First, steady-state circulations driven by localized equatorially symmetric heating are examined. Convectively induced vertical-mode coupling is shown to excite additional baroclinic modes beyond those directly forced, substantially modifying the vertical structure of the Walker circulation. Second, the equatorial wave spectrum is analyzed. The resulting eigenmodes comprise mixtures of the classical equatorial wave families, causing the dispersion relations to lose their distinct branch structure and exhibit systematic frequency shifts. These results demonstrate how convectively induced momentum transport systematically modifies both steady tropical circulations and equatorial wave dynamics.

\end{abstract}


\section{Introduction}
\label{sec:headings}

Deep convection is the primary mechanism by which the tropical atmosphere transports heat and momentum across a wide range of spatial scales \citep{emanuel1986,houze1989}. Through the release of latent heat and the accompanying convective circulations, individual deep convective clouds generate localized buoyancy and momentum anomalies that excite large-scale atmospheric motions \citep{hou:04}. Whilst individual clouds typically have horizontal scales $\lesssim 20\,\mathrm{km}$, the aggregate effect of countless convective elements gives rise to planetary-scale phenomena, including equatorial waves, the Walker circulation, and intraseasonal variability such as the Madden-Julian Oscillation \citep{matsuno:1966,gill:1980,zhang2005,jiang2020}. Understanding how buoyancy and momentum anomalies generated by unresolved convective motions influence the large-scale tropical circulation therefore remains one of the central challenges in theoretical atmospheric dynamics.

The theoretical foundations for understanding the large-scale tropical circulation were established by \cite{matsuno:1966} and \cite{gill:1980}, whose collective contributions have come to be known as the Matsuno--Gill framework, and continue to underpin a wide range of theoretical studies \citep{maher2019}. Matsuno demonstrated that the linearized equatorial $\beta$-plane supports a discrete family of equatorially trapped wave modes, providing the dynamical framework for describing large-scale tropical motions. Gill later incorporated localized diabatic heating to obtain analytical steady-state solutions describing the tropical response to large-scale convective anomalies. In both works, the stratified atmosphere was presumed to decompose into decoupled baroclinic modes, each of which obeys an independent shallow water system with a mode-dependent equivalent depth \citep{gill:1982}. This formulation has provided a mathematically tractable basis for understanding a broad range of tropical atmospheric phenomena, including the Walker circulation \citep{gill:1980,webster1972}, monsoon circulations \citep{rodwell1996,zhang1996}, atmosphere-ocean interactions associated with El Ni\~no \citep{cane1986,neelin1991}, and tropical-extratropical teleconnections \citep{hoskins1981,kucharski2009}. Despite its remarkable success, however, the representation of unresolved convection within the classical Matsuno--Gill framework remains largely phenomenological. In particular, the physical interpretation of the effective damping introduced by Gill has been the subject of considerable discussion \citep{neelin:1989}, since it is prescribed empirically rather than emerging from the underlying convective dynamics. This naturally raises the question of whether the influence of unresolved convection can instead be derived systematically from the underlying multiscale dynamics, thereby providing a first-principles representation of the convective feedback within the large-scale tropical circulation.

Many atmospheric models represent unresolved convection through parameterization schemes for subgrid-scale convective processes \citep{arakawa:1974,betts1986,tiedtke1989,emanuel1991}. Whilst these schemes have proven indispensable for numerical weather prediction and climate modeling, their complexity makes them ill-suited to simplified theoretical descriptions of large-scale tropical dynamics. Several reduced descriptions have been proposed which aim to capture the interaction between convection and the large-scale tropical circulation. Weak-temperature-gradient (WTG) theory exploits the rapid adjustment of gravity waves to constrain large-scale horizontal temperature gradients \citep{sobel2000,raymond2005}, whilst quasi-equilibrium tropical circulation models exploit simplified representations of convective adjustment to obtain reduced descriptions of the tropical atmosphere \citep{neelin2000}. These approaches have provided considerable theoretical insight into tropical dynamics whilst retaining analytical tractability. More recently, multiscale asymptotic techniques have established a systematic framework for deriving effective large-scale equations directly from the multiscale dynamics of convectively forced atmospheres \citep{maj:03,bie:05,bie:10,bie:13,gol:23,gol02:25}. Here, we extend this approach to derive a first-principles closure for the influence of unresolved convection on the large-scale tropical circulation, thereby providing a systematic derivation of the convective feedback represented phenomenologically by Gill's damping.

In the present work, we investigate a multiscale atmosphere comprising small-scale convective circulations embedded within a slowly varying large-scale equatorial flow. Using a systematic multiple-scales analysis based on the theory of homogenization \citep[see e.g.][]{pavliotis:2008,allaire2012}, we derive closed equations governing the large-scale equatorial circulation. The resulting large-scale equations contain a non-local closure for convective momentum transport derived directly from the underlying multiscale dynamics. The classical Matsuno--Gill model is recovered as a limiting approximation in which this closure reduces to the phenomenological damping introduced by Gill. The full closure predicts steady-state circulations that retain the characteristic Kelvin--Rossby structure of the classical Gill solution, whilst convectively induced coupling between vertical modes excites additional baroclinic responses beyond the mode that is directly forced. Consequently, the associated Walker cell acquires a systematic westward tilt with height, suggesting that convectively induced vertical-mode coupling may contribute to the observed vertical structure of the tropical circulation. Furthermore, convectively induced vertical-mode coupling fundamentally modifies the equatorial wave spectrum, preventing the classical wave families from separating into distinct branches and producing systematic frequency shifts in the corresponding dispersion relations. The resulting theory therefore provides a first-principles generalization of the classical Matsuno-Gill framework whilst revealing new dynamical consequences of convectively induced vertical-mode coupling.

The remainder of the paper is organized as follows. Section~\ref{sec:model} derives the asymptotic model, while section~\ref{sec:closure} computes the associated diffusion kernel and coupling matrix for a representative cloud field. Steady-state solutions in response to large-scale convective forcing are examined in section~\ref{sec:steady}, and the corresponding equatorial wave spectrum is analyzed in section~\ref{sec:waves}. Finally, concluding remarks are presented in section~\ref{sec:conclusions}.

\section{Multiscale theory of the tropical circulation}\label{sec:model}
We begin with the nonlinear, rotating Boussinesq equations on an equatorial $\beta$-plane \citep[e.g.][]{gill:1982,vallisbook},
\begin{subequations}
	\label{eq:startingeqns}
	\begin{align}
		\bs{v}_t+\beta y\zhat\times\bs{v}+\bs{v}\cdot\nabla\bs{v}&=-\nabla p+b\zhat-\lambda_m\bs{v},\label{eq:startingeqsmom}\\
		\nabla\cdot\bs{v}&=0,\\
		b_t+\bs{v}\cdot\nabla b+N^2w&=S-\lambda_b b\label{eq:startingeqsbuoy},
	\end{align}
\end{subequations}
where $\bs{v}=(u,v,w)^T$ is the velocity field, with $u$, $v$, and $w$ denoting the zonal, meridional, and vertical velocity components, respectively. The variables $p$ and $b$ denote the pressure and buoyancy perturbations. All variables are presumed to depend on $\bs{x}=(x,y,z)^T$, i.e. the zonal, meridional and vertical coordinates, and time $t$. The parameters featured in equations (\ref{eq:startingeqns}) are the meridional gradient of the planetary vorticity $\beta\approx 2.3\times 10^{-11}\mathrm{m}^{-1}\mathrm{s}^{-1}$ and the \BVfreq~ $N\approx 0.01-0.02\;\mathrm{s}^{-1}$, whilst the coefficients $\lambda_m$ and $\lambda_b$ represent linear momentum and buoyancy damping rates respectively. The buoyancy source term $S$ parameterizes diabatic processes that generate buoyancy anomalies such as latent heat release associated with condensation. 


Our goal is to understand the multiscale interaction between steady cloud-like convective circulations with horizontal scale $L_c$ and equatorial waves with horizontal scale $L_e$ in the distinguished limit $\eps\equiv L_c/L_e\ll1$. Following classical equatorial wave theory \citep[see e.g.][Ch. 11]{gill:1982}, we take the characteristic velocity scale to be the phase speed $c\approx 50\;\mathrm{ms}^{-1}$ of the first baroclinic gravity wave, and the equatorial length-scale as $L_e=\sqrt{c/\beta}\approx 1,500\;\mathrm{km}$. Since the first baroclinic mode has vertical wavenumber $\pi/H$ where $H\approx 15\;\mathrm{km}$ is the tropospheric depth, it is natural to define $L_c=H/\pi$ which yields the phase speed relation $c=NL_c$ and sets $\eps\sim3.3\times 10^{-3}$. We assume that the small-scale convective circulations adjust rapidly relative to the equatorial time-scale and may therefore be treated as quasi-equilibrated at leading order. The temporal variability retained explicitly in the present theory corresponds to slower modulations of this convective state and of the large-scale circulation, occurring on the equatorial time-scale $T_e=1/\sqrt{\beta c}\approx 8\;\mathrm{hrs}$. We therefore nondimensionalize equations (\ref{eq:startingeqns}) according to the scalings
\begin{gather}
	\bs{x}=L_c\bs{x}^*,\quad t=T_e t^*,\quad \bs{v}=c\bs{v}^*,\quad p=c^2p^*,\quad b=c^2/\L_cb^*, \quad S=S_0 S^*,
\end{gather} 
and arrive at the non-dimensional system
\begin{subequations}\label{eq:ndsystem}
	\begin{align}
		\eps \bs{v}_t+\eps^2 y\zhat\times\bs{v}+\bs{v}\cdot\nabla\bs{v}&=-\nabla p+b\zhat-\lambda^*_m\bs{v},\\
		\nabla\cdot\bs{v}&=0,\\
		\eps b_t+\bs{v}\cdot\nabla b+w&=S_0^*S-\lambda^*_b b,		
	\end{align}
\end{subequations}
where for simplicity we have dropped the asterisk notation on all but the dimensionless parameters $\lambda^*_m$, $\lambda^*_b$ and $S^*_0$, which are given by
\begin{gather}\label{eq:distinguishedlimits}
	\lambda^*_m=\frac{\lambda_m L_c}{c},\qquad \lambda^*_b=\frac{\lambda_b L_c}{c},\qquad S^*_0=\frac{S_0}{N^2L_c}.
\end{gather}
To retain momentum damping in the leading-order convective dynamics, while allowing buoyancy damping and residual diabatic forcing to enter the large-scale balance, we adopt the distinguished limits $\lambda_m^*=\eps^\frac{1}{2}\gamma_m$, $\lambda_b^*=\eps\gamma_b$, and $S_0^*=\eps^\frac{1}{2}$, where $\gamma_m,\gamma_b$ are $O(1)$ quantities.

We perform a multiple-scales analysis on equations (\ref{eq:ndsystem}) by first introducing large-scale horizontal spatial coordinates $\bs{X}=(X,Y,0)^T$, where $X=\eps x$ and $Y=\eps y$, and expand the gradient operator using the chain rule as
\begin{equation}\label{eq:gradientexpansion}
	\nabla\to\eps\onab+\nabla,
\end{equation}
where $\onab=(\partial/\partial X,\partial/\partial Y,0)^T$. Upon substituting (\ref{eq:gradientexpansion}) into (\ref{eq:ndsystem}), we find the multiscale system
\begin{subequations}\label{eq:msequations}
	\begin{align}
		\eps\left(\bs{v}_t+Y\zhat\times\bs{v}+\bs{v}\cdot\onab\bs{v}\right)+\bs{v}\cdot\nabla\bs{v}&=-\eps\onab p-\nabla p+b\zhat-\eps^\frac{1}{2}\gamma_m\bs{v},\\
		\eps\onab\cdot\bs{v}+\nabla\cdot\bs{v}&=0,\\
		\eps\left(b_t+\bs{v}\cdot\onab b\right)+\bs{v}\cdot\nabla b+w&=\eps^\frac{1}{2}S-\eps\gamma_b b.
	\end{align}
\end{subequations}

Next, we introduce the averaging operator
\begin{equation}
	\langle\cdots\rangle_\mathcal{C}= \frac{1}{|\mathcal{C}|}\iint_{\mathcal{C}} \cdots\; \mathrm{d}x\mathrm{d}y,
\end{equation}
where $\mathcal{C}$ denotes a horizontal region whose extent is intermediate between the small- and large-scales. This averaging should be interpreted as a local spatial coarse-graining over many unresolved convective elements within $\mathcal{C}$, analogous to the representative averaging cells used in homogenization theory \citep{pavliotis:2008,whitaker:2013}. In the following analysis, we regard the small-scale convective motions as contributing to a coarse-grained field through their aggregate effect within each averaging region, rather than as a superposition of a finite number of discrete, isolated circulations. Accordingly, for a convectively generated field variable $g$, the averaged quantity $\langle g\rangle_\mathcal{C}$ represents a local mean conditioned on the spatial coverage of convection within $\mathcal{C}$. 

Motivated by this interpretation, we suppose that $g$ may be decomposed into a linear superposition of localized contributions distributed across the domain as
\begin{equation}
	g(\bs{x},z,t)=\sum_ig_i(\bs{x}-\bs{x}_i,z,t),
\end{equation}
where each $g_i$ represents a localized convective response centered at $\bs{x}_i$. Introducing a canonical kernel $\hat{g}$ and an occupancy factor $\sigma_i(t)$, which weights the contribution of the $i$th convective element to the coarse-grained convective field, we assume each contribution may be written in the form
\begin{equation}
	g_i(\bs{x}-\bs{x}_i,z,t)=\sigma_i(t)\hat{g}(\bs{x}-\bs{x}_i,z),
\end{equation}
where $\hat{g}$ decays as $|\bs{x}-\bs{x}_i|\to\infty$. Averaging over $\mathcal{C}$, we obtain
\begin{equation}
	\langle g\rangle_\mathcal{C}=\frac{1}{|\mathcal{C}|}\sum_i\sigma_i(t)\iint_\mathcal{C}\hat{g}(\bs{x}-\bs{x}_i,z)\;\mathrm{d}x\mathrm{d}y.
\end{equation}
We assume that $\mathcal{C}$ contains many such localized contributions, and that the distribution of convective elements varies only on the large-scales $(\bs{X},t)$. Thus, after
approximating each localized contribution by its full horizontal integral and replacing the local distribution of convective elements by its coarse-grained average, we arrive at the separable form 
\begin{equation}
	\langle g\rangle_\mathcal{C}(\bs{X},z,t)=M(\bs{X},t)\langle \hat{g}\rangle(z).
\end{equation}
Here, $M(\bs{X},t)$ denotes the \textit{filling fraction}, defined as the fraction of $\mathcal{C}$ occupied by active convective elements, and is given by
\begin{equation}
	M(\bs{X},t)=\frac{A_c}{|\mathcal{C}(\bs{X})|}\sum_{\bs{x}_i\in\mathcal{C}(\bs{X})}\sigma_i(t),
\end{equation} where $A_c$ represents the characteristic nondimensional horizontal area of a localized convective contribution. The quantity $\langle \hat{g}\rangle(z)$, representing the area-normalized mean response associated with a localized convective element, is defined by
\begin{equation}
	\langle\hat{g}\rangle(z)=\frac{1}{A_c}\iint_{\mathds{R}^2}\hat{g}(\bs{\xi},z,t)\;\mathrm{d}\bs{\xi},
\end{equation}
and may be interpreted as the effective vertical structure induced by a unit filling fraction of convection.
In order that contributions from convection appear at leading order in the asymptotics which follow, we set $M(\bs{X},t)=\eps^\frac{1}{2}\mu(\bs{X},t)$, where $\mu$ is of order $O(1)$.

We decompose the fields in (\ref{eq:msequations}) into their slowly-varying averages (denoted by uppercase variables) and their small-scale zero-average fluctuations (expanded in powers of $\eps^{1/2}$) as
\begin{subequations}\label{eq:perturbationexpansions}
	\begin{align}
		\bs{v}(\bs{X},\bs{x},t)&=\eps \bs{U}(\bs{X},z,t)+\eps^2 W(\bs{X},z,t)\zhat+\eps^\frac{1}{2}\sum_{i=0}^\infty\eps^\frac{i}{2}\bs{v}_i(\bs{X},\bs{x},t),\\
		p(\bs{X},\bs{x},t)&=\eps P(\bs{X},z,t)+\eps\sum_{i=0}^\infty\eps^\frac{i}{2}p_i(\bs{X},\bs{x},t),\\
		b(\bs{X},\bs{x},t)&=\eps B(\bs{X},z,t)+\eps\sum_{i=0}^\infty\eps^\frac{i}{2}b_i(\bs{X},\bs{x},t).
	\end{align}
\end{subequations}	
Here, $\bs{U}=(U,V,0)^T$ denotes the slowly-varying averaged horizontal velocity field, whilst $W$ represents the corresponding vertical component. The prefactors in each term are chosen so that the leading-order convectively generated velocity perturbations have dimensional magnitude $O(\eps^{1/2}c)$, corresponding to several meters per second,
whilst the large-scale horizontal velocity field is asymptotically weaker by a factor of $\eps^{1/2}$. We further decompose the diabatic source term according to
\begin{equation}\label{eq:heatdecomposition}
	S(\bs{x})=S_0(\bs{x})+\eps S_1(\bs{x}),
\end{equation}
where $S_0$ denotes the leading-order horizontally localized buoyancy source responsible for driving the small-scale convective circulations, and $S_1$ parameterizes a weaker component describing the leakage of buoyancy production into larger scales. 

\subsection{Leading-order convective circulation}
Upon substituting the expansions (\ref{eq:perturbationexpansions}) and (\ref{eq:heatdecomposition}) into (\ref{eq:msequations}) and gathering terms at leading order (i.e. $O(\eps^{1/2})$ in the continuity and buoyancy equations, and $O(\eps)$ in the momentum equation), we find
\begin{subequations}\label{eq:leadingordereqs}
	\begin{align}
		\bs{v}_0\cdot\nabla\bs{v}_0&=-\nabla p_0+b_0\zhat-\gamma_m\bs{v}_0,\label{eq:loeqa}\\
		\nabla\cdot\bs{v}_0&=0,\\
		w_0&=S_0.
	\end{align}
\end{subequations}
Equations (\ref{eq:leadingordereqs}) describe the leading-order small-scale response of the atmosphere to a diabatic buoyancy source under the \textit{convective weak temperature gradient} (C-WTG) approximation \citep[][]{mar01:23} -- a local analog of the WTG regime commonly employed over large-scale tropical regions \citep[see e.g.][]{sobel2000,raymond2005}. A notable consequence of the C-WTG framework is the diagnostic relationship $w_0=S_0$, which determines the leading-order vertical velocity directly from the prescribed buoyancy source. Thus, specifying the form of $S_0$ is equivalent to prescribing the leading-order vertical velocity field $w_0$. 

Consistent with the coarse-graining framework above, we focus on the case in which the leading-order forcing may be regarded as a superposition of canonical localized axisymmetric sources of the form $\hat{S}_0(r,z)$, each associated with an individual convective element. The corresponding solutions of (\ref{eq:leadingordereqs}) then define the canonical response kernels for the velocity, pressure and buoyancy fields. In what follows, we restrict attention to a single canonical convective element and its associated source $\hat{S}_0$. For notational simplicity, the hat notation will be suppressed; henceforth, all subscripted variables are understood to denote successive terms in the asymptotic expansion of the canonical convective response. In axisymmetric coordinates, the leading-order velocity field takes the form 
\begin{equation}\label{eq:leadingordervelocityfield}
	\bs{v}_0(r,z)=u_0^r(r,z)\rhat+w_0(r,z)\zhat,
\end{equation}
so that the leading-order flow constitutes a \textit{poloidal} circulation \citep{gol01:25,gol02:25,mar01:23}.

The radial component of the velocity $u_0^r$ is determined from the vertical velocity via the integral relation
\begin{equation}
	u_0^r(r,z)=-\frac{1}{r}\int_0^r\partial_zw_0(r',z)\;r'\mathrm{d}r',
\end{equation}
whilst the leading-order buoyancy $b_0$ is determined from the azimuthal component of the vorticity equation at leading order. Taking the curl of (\ref{eq:loeqa}), we obtain
\begin{equation}
	\bs{v}_0\cdot\nabla\omega_0-\frac{\omega_0u_0^r}{r}=-\partial_rb_0,
\end{equation}
where $\omega_0=\partial_zu_0^r-\partial_rw_0$ is the azimuthal component of the leading-order vorticity, and thus we find
\begin{equation}\label{eq:buoyancyintegral}
	b_0(r,z)=\int_r^\infty\bs{v}_0(r',z)\cdot\nabla\omega_0(r',z)-\frac{\omega_0(r',z)u^r_0(r',z)}{r'}\;\mathrm{d}r'.
\end{equation}
For completeness, the leading-order pressure field $p_0$ is found via the Leray projection
\begin{equation}
	p_0=\nabla^{-2}\left(\partial_z b_0-\nabla\cdot(\bs{v}_0\cdot\nabla\bs{v}_0)\right),
\end{equation}
where the Laplacian operator is inverted subject to appropriate boundary conditions; however, for our purposes we are not required to compute $p_0$.

\subsection{First-order horizontal recirculation}
The buoyancy equation at $O(\eps)$ yields $w_1=0$, implying that the first-order correction to the velocity field $\bs{v}_1$ takes the form
\begin{equation}
	\bs{v}_1(\bs{X},\bs{x},t)=u_1^r(\bs{X},\bs{x},t)\rhat+u_1^\theta(\bs{X},\bs{x},t)\that,
\end{equation}
  lying entirely within the horizontal plane. Consequently, using axisymmetric coordinates, we describe $\bs{v}_1$ using a streamfunction $\psi_1$ defined through the relations
\begin{equation}\label{eq:o1streamfunction}
	u_1^r=\frac{1}{r}\frac{\partial\psi_1}{\partial \theta},\qquad u_1^\theta=-\frac{\partial\psi_1}{\partial r}.
\end{equation}
An equation specifying $\psi_1$ emerges from Ertel's potential vorticity equation (see appendix~\ref{app:ErtelsPV}), which at leading order is found to be
\begin{equation}\label{eq:cellprobem}
	\left(u_0^r\partial_r+w_0\partial_z\right)\nabla^2_h\psi_1+\partial_rw_0\partial^2_{rz}\psi_1-\partial_zw_0\nabla^2_h\psi_1+\gamma_m\nabla^2_h\psi_1=(\that\cdot\partial_z\bs{U})\partial_rw_0-Y\partial_zw_0.
\end{equation}	
where $\nabla^2_h=\partial^2_{rr}+r^{-1}\partial_r+r^{-2}\partial^2_{\theta\theta}$ denotes the horizontal Laplacian.

Equation (\ref{eq:cellprobem}), termed the \textit{cell problem}, couples the small-scale velocity corrections associated with an individual convective element to the large-scale vorticity field. It is strongly reminiscent of the cell problem identified by \cite{gol02:25} in the context of wind shear forcing by the non-traditional Coriolis terms, and as such, many of the following analytical forms are identical to theirs. The cell problem may be interpreted as a linear vorticity equation governing the horizontal recirculation $\psi_1$. The recirculation is advected, tilted and stretched by the leading-order poloidal circulation, whilst being forced by two distinct mechanisms: the tilting of environmental vertical shear by the convective updraft, and the stretching of planetary vorticity by the circulation itself. Following \cite{gol01:25,gol02:25}, a solution to (\ref{eq:cellprobem}) is sought in the form
\begin{equation}\label{eq:recirculationresponse}
	\psi_1(\bs{X},\bs{x},t)=\int_0^\pi \psihat(r,z,z')\that\cdot\partial_{z'}\bs{U}(\bs{X},z',t)\;\mathrm{d}z'+\Pi(r,z)Y,
\end{equation}
where $\psihat(r,z,z')$ is the response kernel describing the recirculation generated by a unit impulse in the environmental shear at height $z'$. Subtituting (\ref{eq:recirculationresponse}) into (\ref{eq:cellprobem}) yields the \textit{kernel cell problem} 
\begin{subequations}\label{eq:kernelcellproblem}
	\begin{align}
		\left(u_0^r\partial_r+w_0\partial_z\right)\zetahat-\partial_rw_0\partial^2_{rz}\psihat-\partial_zw_0\zetahat+\gamma_m\zetahat&=-\delta(z-z')\partial_rw_0,\\
		\zetahat&=-\partial^2_{rr}\psihat-\frac{1}{r}\partial_r\psihat+\frac{\psihat}{r^2},
	\end{align}
\end{subequations}
alongside a local PDE governing the response to planetary vorticity forcing through the function $\Pi(r,z)$. The component of $\psi_1$ induced by the planetary vorticity gradient is not of primary importance for this work, and thus further discussion of $\Pi(r,z)$ is relegated to appendix~\ref{app:vorticitycorrection}.

\subsection{Effective large-scale equations}
Applying the averaging operator to equations (\ref{eq:msequations}) and gathering terms at leading order\footnote{It should be noted that the averaged horizontal momentum equation produces an apparently larger term $\gamma_m\bs{U}$, which appears at $O(\eps^{3/2})$. This term is presumed to be balanced by large-scale atmospheric processes lying outside the scope of the present model and is therefore omitted from the reduced equations.} (i.e. $O(\eps^2)$ in the horizontal momentum, the continuity and the buoyancy equations, and $O(\eps)$ in the vertical momentum equation), we find
\begin{subequations}\label{eq:averagedeqs}
\begin{align}
	\bs{U}_t+Y\zhat\times\bs{U}&=-\onab P-\mu\partial_z\langle w_0\bs{v}_1\rangle,\\
	P_z&=B,\\
	\onab\cdot\bs{U}+W_z&=0,\\
	B_t+W&=-\gamma_b B+\mu F(z),
\end{align}
\end{subequations}
where
\begin{equation}
	F(z)=\langle S_1\rangle -\partial_z\langle w_0b_0\rangle,
\end{equation}
is the net convective heating of the large-scale flow. It combines the horizontally averaged residual heating with the convergence of the leading-order convective buoyancy flux.

Equations (\ref{eq:averagedeqs}) therefore form a linear, hydrostatic system for the large-scale circulation, subject to radiative damping, and modified by two effective convective contributions. The term $\mu F(z)$ represents the net thermal forcing supplied by the small scales. Although its buoyancy-flux convergence contribution can be diagnosed from the leading-order convective circulation, the residual heating $\langle S_1\rangle$ is not determined by the present theory, and only their sum enters the large-scale equations. Henceforth we treat $F(z)$ as a prescribed quantity. In principle, a more complete model incorporating cloud microphysics and thermodynamic processes could determine the residual heating and thereby provide a systematic closure for $F(z)$. 

By contrast, the momentum-flux convergence $-\mu\partial_z\langle w_0\bs{v}_1\rangle$ constitutes a dynamical feedback on the large-scale circulation. Because $\bs{v}_1$ is the response of the small-scale convective flow to $\bs{U}$, this term must be closed in terms of the large-scale horizontal velocity field. 
Substituting (\ref{eq:recirculationresponse}), we find
\begin{equation}\label{eq:cumulusdrag}
	-\mu\partial_z\langle w_0\bs{v}_1\rangle=\mu\partial_z(\mathcal{K}\partial_z\bs{U}),
\end{equation}
where the operator $\mathcal{K}$ acts on an arbitrary function $f(z)$ as
\begin{equation}
	\mathcal{K}f(z)=\int_0^\pi K(z,z')f(z')\;\mathrm{d}z',
\end{equation}
and the diffusion kernel $K(z,z')$ is defined by
\begin{equation}\label{eq:kernel}
	K(z,z')=\frac{\pi}{A_c}\int_0^\infty w_0(r,z)\partial_r(r\psihat(r,z,z'))\;\mathrm{d}r.
\end{equation}
The kernel $K(z,z')\geq0$ is non-negative (see \S~\ref{sec:closure} below).
Thus, the momentum flux convergence reduces to a non-local vertical diffusion operator acting on the large-scale horizontal velocity, and the effective diffusion is determined by the structure of the convective field. The key result is the derivation of a first-principles parameterization of convective momentum transport, obtained directly from the small-scale convective dynamics rather than introduced phenomenologically.

\subsection{Baroclinic mode decomposition}
For our purposes, the most practical approach to analyzing steady-state solutions and wave dynamics involves decomposing (\ref{eq:averagedeqs}) into baroclinic modes. We expand the large-scale averaged variables in Fourier sine and cosine series as
\begin{subequations}
\begin{align}
	[U,V,P](\bs{X},z,t)&=\sum_{m=1}^{N_m}[U_m,V_m,P_m](\bs{X},t)\cos(mz),\\ [W,B](\bs{X},z,t)&=\sum_{m=1}^{N_m}[W_m,B_m](\bs{X},t)\sin(mz),
\end{align}
\end{subequations}
truncated at the $N_m$th vertical mode.

Firstly, we see that the solution to the cell problem (\ref{eq:cellprobem}) may be written as
\begin{equation}
	\psi_1(\bs{X},\bs{x},t)=\sum_{n=1}^{N_m}\psitil_n(r,z)\left(\that\cdot\bs{U}_n(\bs{X},t)\right), 
\end{equation}
where the $\psitil_n$'s satisfy the \textit{modal cell problem} given by
\begin{subequations}\label{eq:modalcellproblem}
	\begin{align}
		\left(u_0^r\partial_r+w_0\partial_z\right)\zetatil_n-\partial_rw_0\partial^2_{rz}\psitil_n-\partial_zw_0\zetatil_n+\gamma_m\zetatil_n&=n\sin(nz)\partial_rw_0,\\
		\zetatil_n&=-\partial^2_{rr}\psitil_n-\frac{1}{r}\partial_r\psitil_n+\frac{\psitil_n}{r^2}.
	\end{align}
\end{subequations}
The modal cell problem is significantly more tractable than the kernel cell problem from a numerical standpoint due to the smoothness of the right-hand side, and in principle it may be solved for any number of baroclinic modes. 
The averaged equations then become
\begin{subequations}\label{eq:coupledaveragedeqs}
	\begin{align}
		\partial_t\bs{U}_m-Y\zhat\times\bs{U}_m&=-\onab P_m+\mu \sum_{n=1}^{N_m} K_{mn}\bs{U}_n,\\
		\partial_tP_m+\frac{1}{m^2}\onab\cdot\bs{U}_m&=-\gamma_b P_m -\frac{\mu}{m}F_m,
	\end{align}	
\end{subequations}
where the coupling matrix components $K_{mn}$ are defined by 
\begin{equation}\label{eq:couplingmatrix}
	K_{mn}=\frac{2m}{A_c}\int_0^\pi\int_0^\infty w_0\partial_r(r\psitil_n)\sin(mz)\;\mathrm{d}r\mathrm{d}z,
\end{equation}
and
\begin{equation}
	F_m=\frac{2}{\pi}\int_0^\pi F(z)\sin(mz)\;\mathrm{d}z,
\end{equation}
are the Fourier sine coefficients of the thermal forcing.

Equations (\ref{eq:coupledaveragedeqs}), which represent the vertical decomposition into baroclinic modes of equations (\ref{eq:averagedeqs}), provide an alternative interpretation of the effects of small-scale convection on the large-scale momentum balance. Rather than appearing as a direct vertical diffusion operator, the convective momentum transport manifests itself as a coupling between baroclinic modes through an $N_m\times N_m$ matrix $\mathbf{K}$ whose components are $K_{mn}$. The diagonal components $K_{mm}$ describe the damping or amplification of the $m$th baroclinic mode as a result of convection, whilst off-diagonal entries quantify the transfer of momentum between distinct vertical structures. In the absence of off-diagonal terms, each baroclinic mode evolves independently according to a damped shallow water system with its own equivalent depth, analogous to the classical Matsuno--Gill model formulations. The present theory generalizes this picture by showing that unresolved convection naturally couples baroclinic modes. Consequently, even if the large-scale flow initially occupies only a single baroclinic mode, interaction with the unresolved convective field will generally excite additional vertical structures at later times.

\section{The convective closure}\label{sec:closure}
In this section we prescribe a representative convective element and compute the diffusion kernel and coupling matrix required to close the large-scale equations derived in the previous section. Since the leading-order equations (\ref{eq:leadingordereqs}), kernel cell problem (\ref{eq:kernelcellproblem}) and diffusion kernel (\ref{eq:kernel}) are identical to those found by \cite{gol02:25}, we restrict attention to a brief summary of the key results. The novelty of this work lies not in the small-scale structures themselves, but rather in the manner in which they affect large-scale tropical dynamics through the asymptotic closure above.

\begin{figure}
	\centering
	\includegraphics[width=\textwidth]{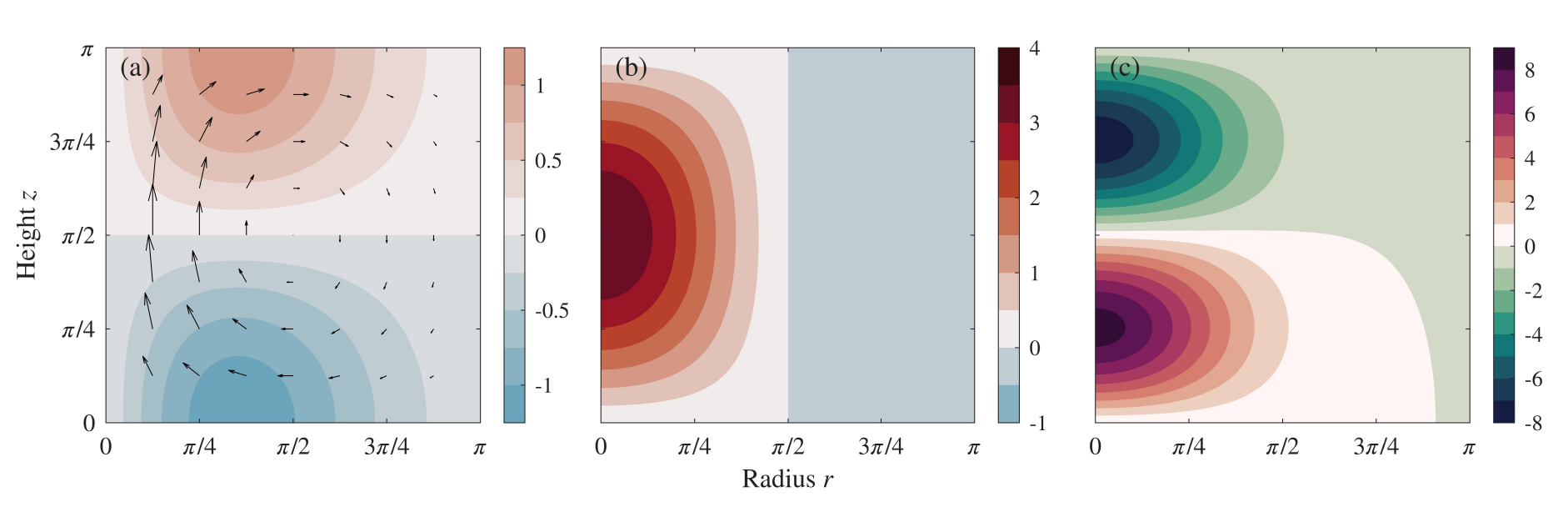}
	\caption{Leading-order fields associated with the prescribed convective circulation: Panel (a) shows the radial velocity $u_0^r(r,z)$, panel (b) shows the vertical velocity $w_0(r,z)$, and panel (c) shows the buoyancy perturbation $b_0(r,z)$. Panel (a) additionally contains vectors showing the relative magnitude and direction of the circulation. The parameters used are $W_0=3.5$ and $\gamma_m=0.1$.}
	\label{fig:flowfields}
\end{figure}
 
We prescribe the leading order velocity components $u_0^r$ and $w_0$ (equivalent to specifying the diabatic buoyancy source $S_0$ in the C-WTG regime) as
\begin{subequations}\label{eq:loflow}
	\begin{align}
		u^r_0(r,z)&=-r\frac{W_0}{2}\exp(-r^2/R^2)\cos(z),\\
		w_0(r,z)&=W_0(1-r^2/R^2)\exp(-r^2/R^2)\sin(z),
	\end{align}
\end{subequations}
corresponding to an idealized deep-convective element extending through the full depth of the troposphere, with updraft radius $R$ and maximum vertical velocity $W_0$. Fig.~\ref{fig:flowfields} shows contours of the prescribed velocity fields $u_0^r(r,z)$ and $w_0(r,z)$, as well as the buoyancy perturbation field $b_0(r,z)$ computed from (\ref{eq:buoyancyintegral}). The resulting circulation bears the qualitative characteristics of a deep convective cell, with a strong vertical motion through the updraft region $r<R$ and weaker converging/diverging horizontal winds near the surface/tropopause. The parameter $W_0=3.5$ is chosen so that the maximum dimensional updraft speed is $\sim 10\;\mathrm{ms}^{-1}$, whilst a modest value of momentum damping $\gamma_m=0.1$ ensures that the diagnosed buoyancy field is positive/negative in the lower/upper troposphere. Following these definitions, we set the convective area to $A_c=\pi R^2$ corresponding to the horizontal area occupied by the updraft.

The kernel equations (\ref{eq:kernelcellproblem}) and modal cell problem (\ref{eq:modalcellproblem}) are each solved on $(r,z)\in[0,\infty)\times[0,\pi]$ using a Chebyshev spectral scheme \citep{trefethenbook,boy:01} with $N_r=150$ and $N_z=52$ radial and vertical collocation points. Regularity is imposed at the symmetry axis $r=0$, homogeneous Neumann boundary conditions are applied at the upper and lower boundaries $z=0,\pi$, and the solutions are required to decay as $r\to\infty$. The semi-infinite radial domain is represented using a conformal mapping \citep[see e.g.][]{boyd1982}. When solving the kernel cell problem, discontinuities which arise as a result of the Dirac delta forcing are dealt with following the methodology of \cite{jun:09}. In order to improve numerical stability, a small eddy viscosity term with coefficient $\nu=10^{-3}$ is included.

\begin{figure}
	\centering
	\includegraphics[width=\textwidth]{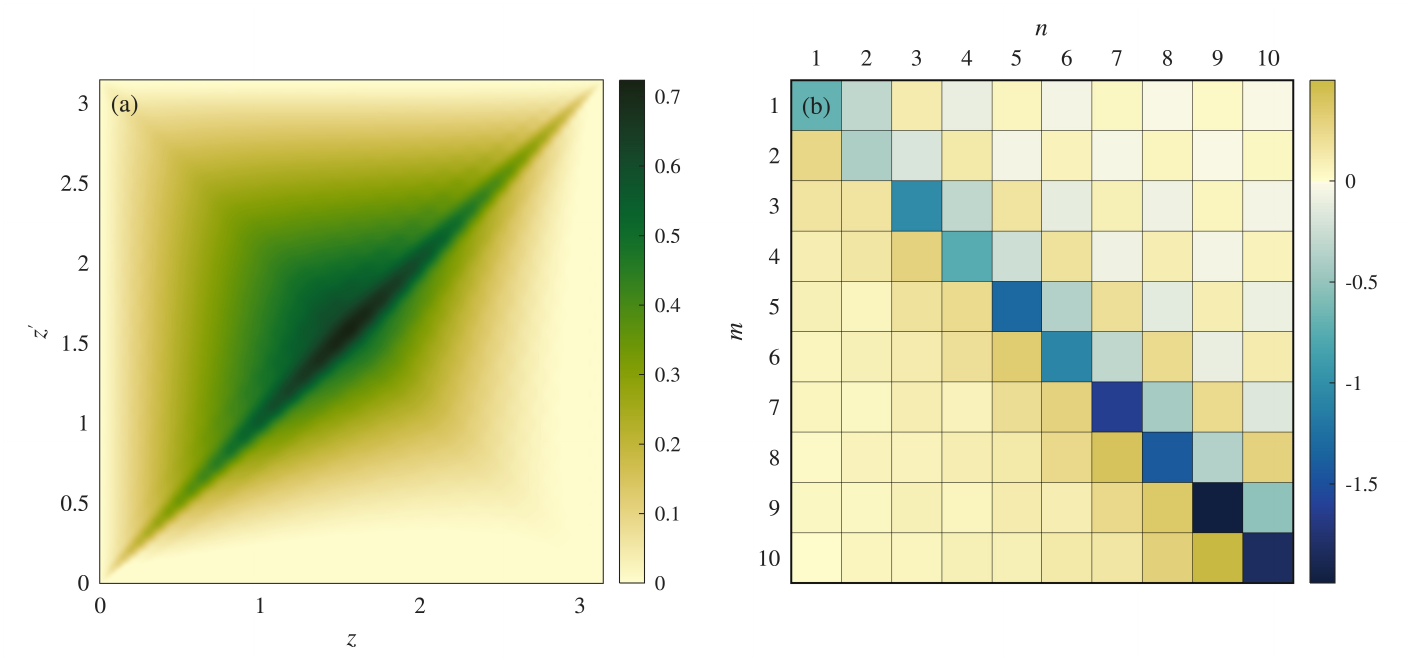}
	\caption{Panel (a): Diffusion kernel $K(z,z')$ computed from (\ref{eq:kernel}) and (\ref{eq:kernelcellproblem}). Panel (b): Coupling matrix $K_{mn}$ computed from (\ref{eq:couplingmatrix}) and (\ref{eq:modalcellproblem}) with $N_m=10$. The convective area is taken to be $A_c=\pi R^2$. In both panels, the quantities are computed from the leading-order flow field (\ref{eq:loflow}) using parameter values $R=\pi/2$, $W_0=3.5$, $\gamma_m=\gamma_b=0.1$, and $\nu=10^{-3}$. }
	\label{fig:kernelmatrix}
\end{figure}

The diffusion kernel $K(z,z')$ shown in fig.~\ref{fig:kernelmatrix}a is positive and concentrated around $z=z'$, indicating that the primary effect of convection is to redistribute horizontal momentum between neighbouring vertical levels. Nevertheless, its appreciable width away from the diagonal demonstrates that this redistribution is genuinely nonlocal, with the strongest coupling occurring through the middle troposphere and weakening near the upper and lower boundaries. This structure is similarly reflected in the modal coupling matrix $K_{mn}$ shown in fig.~\ref{fig:kernelmatrix}b. The matrix is dominated by negative diagonal entries with increasing magnitude at higher vertical-mode numbers, consistent with increasingly strong diffusive damping of modes having shorter vertical scales. The weaker, but nonzero, off-diagonal entries transfer momentum between distinct vertical modes, and therefore prevent the coupled system from being interpreted as a collection of independently damped baroclinic modes. Moreover, $K_{mn}$ is not symmetric, meaning that its off-diagonal contribution may modify wave propagation dispersively as well as produce decay. These properties motivate the examination of the fully coupled steady-state and wave responses in the following sections.

\section{Steady-state solutions}\label{sec:steady}
In this section we derive steady-state solutions of (\ref{eq:coupledaveragedeqs}) under the equatorial long-wave scaling for prescribed large-scale buoyancy forcing. The steady problem is intended as an idealized representation of the large-scale tropical response induced by a geographically localized enhancement of convective activity. We therefore perturb the spatially uniform radiative-convective equilibrium by prescribing a localized variation in the convective filling fraction. This provides the direct analogue, within the present framework, of the localized diabatic heating used to force the classical Matsuno-Gill Problem.

 The linear equatorial long-wave equations emerge from the distinguished limit $L_Y/L_X\ll1$, where $L_X$ and $L_Y$ represent the zonal and meridional length-scales associated with the large-scale flow \citep{matsuno:1966,gill:1980,majda2003}. In this limit, the meridional momentum equation reduces to geostrophic balance, whilst the zonal momentum equation remains prognostic. Accordingly, we decompose the filling fraction as $\mu=\overline{\mu}+\mu'(\bs{X})$ where $\overline{\mu}$ is spatially uniform, and $\mu'$ denotes perturbations, and linearize about the radiative-convective equilibrium satisfying $\gamma_bP_m=\overline{\mu}F_m/m$ and $U_m=V_m=0$ for all $m=1,\dots,N_m$.
The perturbation equations then become
\begin{subequations}\label{eq:firstbcmodesteady}
	\begin{align}
		-YV_m&=-\partial_XP_m+\overline{\mu}\sum_{n=1}^{N_m} K_{mn}U_n,\\
		YU_m&=-\partial_YP_m,\\
		\gamma_bP_m+\tfrac{1}{m^2}\left(\partial_XU_m+\partial_YV_m\right)&=-\tfrac{1}{m}\mu'F_m,
	\end{align}
\end{subequations}
where we reuse $U_m,V_m,P_m$ to represent perturbations about the basic state. For the remainder of this section, we will consider an equatorially symmetric forcing profile for which the filling fraction perturbation is taken to be 
\begin{equation}\label{eq:fillingfrctionform}
	\mu'(X,Y)=\phi_0(Y)\mu_0(X),
\end{equation} 
where $\phi_0(Y)=\pi^{-\frac{1}{4}}\exp(-Y^2/2)$ is the zeroth-order parabolic cylinder function and $\mu_0(X)$ is an arbitrary function which decays as $X\to\pm\infty$.
We adopt the forcing profile (\ref{eq:fillingfrctionform}) because its meridional structure admits an exact decomposition in the parabolic cylinder function basis. As shown below, in the classical uncoupled limit the response is confined to the first three parabolic cylinder functions \citep[see e.g.][]{majda2003}, providing a useful point of comparison with the coupled system.

The primary purpose of this section is to establish a direct connection between the present theory and the classical Matsuno--Gill model. In particular, we show that the phenomenological damping introduced by \cite{gill:1980} emerges naturally from the convective momentum transport parameterization derived above, thereby providing a systematic physical interpretation of heuristics employed in classical theory. Having established this connection, we then investigate how the additional vertical mode coupling predicted by the homogenization procedure modifies the classical steady response.

\subsection{Classical theory: analytical solutions}
In Gill's classical theory of steady equatorial circulations, linear momentum and buoyancy damping are retained, whereas coupling between baroclinic modes is neglected. Furthermore, for simplicity, the momentum and buoyancy damping coefficients are taken to be equal. Within the present framework, this corresponds to retaining only the diagonal entries of $\mathbf{K}$ and equating $\gamma_b$ with the resulting momentum damping coefficient for the relevant baroclinic mode in (\ref{eq:firstbcmodesteady}). After equating the effective modal linear damping coefficients $-\overline{\mu}K_{mm}=m^2\gamma_b\equiv\kappa$, and defining the effective modal forcing coefficient $mF_m\equiv F$, we find equations for the $m$th baroclinic mode  
\begin{subequations}\label{eq:gillmodel}
	\begin{align}
		\kappa U-Y V&=-P_X,\\
		YU&=-P_Y,\\
		\kappa P+U_X+V_Y&=-F\mu'(X,Y), 
	\end{align}
\end{subequations}
where we have dropped the subscript $m$ for notational simplicity.

\begin{figure}
	\centering
	\includegraphics[width=0.9\textwidth]{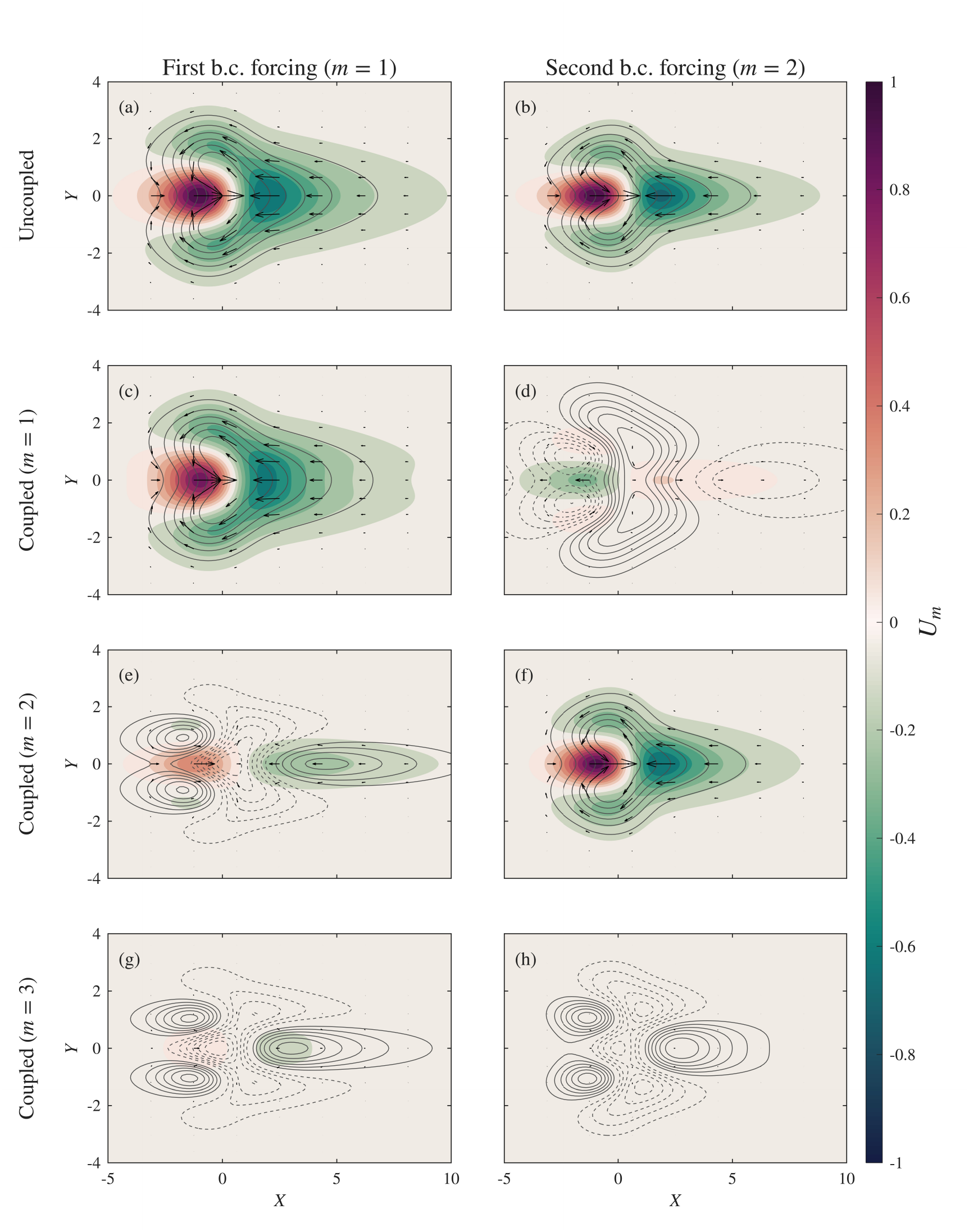}
	\caption{Steady-state solutions to (\ref{eq:firstbcmodesteady}) in response to baroclinic forcing. Left column: forcing applied in the first baroclinic mode. Panel (a) shows the corresponding classical Gill solution given by (\ref{eq:kelvinrossbyerfcsols}) for a Gaussian zonal forcing profile, whilst panels (c), (e), and (g) show the first, second, and third baroclinic responses, respectively, for the equivalent coupled system computed from (\ref{eq:coupledsteadysystemexpansion}). Right column: as in the left column, but for forcing applied in the second baroclinic mode, with the classical Gill solution shown in panel (b) and the first, second, and third baroclinic responses shown in panels (d), (f), and (h), respectively. In each column, the forcing amplitude is chosen so that the maximum zonal velocity is unity, with background filling fraction $\overline{\mu}=1/2$. Colour indicates zonal velocity using common colour levels across all panels. Pressure is shown by contours, with contour intervals chosen independently for each panel; solid contours denote negative pressure and dashed contours denote positive pressure. Arrows indicate the corresponding horizontal circulation. The numerical computations use $N_l=10$ meridional modes and $N_m=15$ baroclinic modes.}
	\label{fig:steadyplot}
\end{figure}

For the particular filling fraction perturbation profile (\ref{eq:fillingfrctionform}), it is well known that  the solutions to (\ref{eq:gillmodel}) may be expressed as linear combinations of the first three parabolic cylinder functions $\{\phi_0(Y),\phi_1(Y),\phi_2(Y)\}$ \citep[see e.g.][]{majda2003,majda2009}. The resulting circulation consists of an equatorially trapped Kelvin-like response together with a pair of off-equatorial gyres whose meridional structure resembles that of the first symmetric Rossby mode. We seek solutions to (\ref{eq:gillmodel}) using the ansatz
\begin{subequations}\label{eq:kelvinrossbyansatz}
	\begin{align}
			U(X,Y)&=[A_\mathrm{K}(X)-A_\mathrm{R}(X)]\phi_0(Y)+\frac{1}{\sqrt{2}}A_\mathrm{R}(X)\phi_2(Y),\\
			V(X,Y)&=\frac{1}{\sqrt{2}}[A'_R(X)+\kappa A_\mathrm{R}(X)]\phi_1(Y),\\
			P(X,Y)&=[A_\mathrm{K}(X)+A_\mathrm{R}(X)]\phi_0(Y)+\frac{1}{\sqrt{2}}A_\mathrm{R}(X)\phi_2(Y),
	\end{align}
\end{subequations} 
where $A_\mathrm{K}$ and $A_\mathrm{R}$ denote the zonal amplitude functions associated with the Kelvin-like and Rossby-like components, respectively. Upon substituting (\ref{eq:kelvinrossbyansatz}) into (\ref{eq:gillmodel}), we find equations for the modal amplitudes
\begin{equation}
	A_\mathrm{K}'+\kappa A_\mathrm{K}=-\tfrac{1}{2}F\mu_0(X),\qquad A_\mathrm{R}'-3\kappa A_\mathrm{R}=F\mu_0(X),
\end{equation}
whose bounded solutions are readily found to be
\begin{equation}\label{eq:kelvinrossbyamplitudes}
	A_\mathrm{K}(X)=-\tfrac{1}{2}F\int_{-\infty}^X\mathrm{e}^{-\kappa(X-\xi)}\mu_0(\xi)\;\mathrm{d}\xi,\qquad A_\mathrm{R}(X)=-F\int_X^\infty \mathrm{e}^{-3\kappa(\xi-X)}\mu_0(\xi)\;\mathrm{d}\xi.
\end{equation}
For the particular case of a Gaussian zonal forcing profile $\mu_0(X)=\exp(-X^2/2)$, the integrals in (\ref{eq:kelvinrossbyamplitudes}) may be evaluated explicitly to give
\begin{subequations}\label{eq:kelvinrossbyerfcsols}
	\begin{align}
		A_\mathrm{K}(X)
		&=-F\sqrt{\frac{\pi}{8}}
		\mathrm{e}^{\kappa^2/2-\kappa X}
		\operatorname{erfc}\left(\frac{\kappa-X}{\sqrt{2}}\right),\\
		\qquad
		A_\mathrm{R}(X)
		&=-F\sqrt{\frac{\pi}{2}}
		\mathrm{e}^{9\kappa^2/2+3\kappa X}
		\operatorname{erfc}\left(\frac{X+3\kappa}{\sqrt{2}}\right),
	\end{align}
\end{subequations}
where $\operatorname{erfc}$ denotes the complementary error function. Steady-state solutions to the classical Gill system are computed using (\ref{eq:kelvinrossbyerfcsols}), and shown in the left-most column of fig.~\ref{fig:steadyplot}.

\subsection{Coupled system: numerical framework}
We now turn our attention to the more general case of solving the fully coupled long-wave system (\ref{eq:firstbcmodesteady}). Applying the Fourier transform $\mathcal{F}$, defined by
\begin{equation}
	\hat{f}(k)=\mathcal{F}[f]:=\int_{-\infty}^\infty f(X)\mathrm{e}^{-\mathrm{i}kX}\;\mathrm{d}X,
\end{equation}
to (\ref{eq:firstbcmodesteady}), we find
\begin{subequations}\label{eq:fouriertransformeqs}
	\begin{align}
		-Y\hat{V}_m&=-\mathrm{i}k\hat{P}_m+\overline{\mu}\sum_{n=1}^\infty K_{mn}\hat{U}_n,\\
		Y\hat{U}_m&=-\partial_Y\hat{P}_m,\\
		\gamma_b\hat{P}_m+\tfrac{1}{m^2}\left(\mathrm{i}k\hat{U}_m+\partial_Y\hat{V}_m\right)&=-\tfrac{1}{m}F_m\hat{\mu}'(k).
	\end{align}
\end{subequations}
We next expand the meridional dependence of the solution in the truncated basis of normalized parabolic cylinder functions, $\{\phi_l(Y)\}_{l=0,\dots,2N_l+1}$ as
\begin{subequations}\label{eq:paraboliccylinderfunctionexpansion}
\begin{align}
	\left[\hat{U}_m,\hat{P}_m\right](k,Y)&=\sum_{l=0}^{N_l}[\hat{U}_{2l,m},\hat{P}_{2l,m}](k)\phi_{2l}(Y),\\ \hat{V}_m(k,Y)&=\sum_{l=0}^{N_l}\hat{V}_{2l+1,m}(k)\phi_{2l+1}(Y),
\end{align}
\end{subequations}
where we have exploited equatorial symmetry to retain only even modes for $\hat{U}_m$ and $\hat{P}_m$ and odd modes for $\hat{V}_m$. The recurrence relations \citep[see e.g.][]{abramowitz1964handbook}
\begin{subequations}\label{eq:pcfidentities}
	\begin{align}
		Y\phi_l(Y)&=\sqrt{\tfrac{l+1}{2}}\phi_{l+1}(Y)+\sqrt{\tfrac{l}{2}}\phi_{l-1}(Y),\\ \phi_l'(Y)&=-\sqrt{\tfrac{l+1}{2}}\phi_{l+1}(Y)+\sqrt{\tfrac{l}{2}}\phi_{l-1}(Y),
	\end{align}
\end{subequations}
allow the transformed equations to be projected onto the truncated parabolic cylinder basis.
Substituting (\ref{eq:paraboliccylinderfunctionexpansion}) into (\ref{eq:fouriertransformeqs}) and projecting onto the $l$th meridional mode yields
\begin{subequations}\label{eq:coupledsteadysystemexpansion}
	\begin{align}
		-\sqrt{l}\hat{V}_{2l-1,m}-\sqrt{l+\tfrac{1}{2}}\hat{V}_{2l+1,m}+\mathrm{i}k\hat{P}_{2l,m}-\overline{\mu}\sum_{n=1}^{N_m} K_{mn}\hat{U}_{2l,m}&=0,\\
		\sqrt{l+\tfrac{1}{2}}\hat{U}_{2l,m}+\sqrt{l+1}\hat{U}_{2l+2,m}-\sqrt{l+\tfrac{1}{2}}\hat{P}_{2l,m}+\sqrt{l+1}\hat{P}_{2l+2,m}&=0,\\
		\gamma_b\hat{P}_{2l,m}+\tfrac{1}{m^2}\Big(\mathrm{i}k\hat{U}_{2l,m}-\sqrt{l}\hat{V}_{2l-1,m}+\sqrt{l+\tfrac{1}{2}}\hat{V}_{2l+1,m}\Big)&=-\tfrac{1}{m}F_m\hat{\mu}_0(k)\delta_{l0},
	\end{align}
\end{subequations}
where $\delta_{l0}$ denotes the Kronecker delta, and where all coefficients with indices outside of the retained range are taken to vanish, i.e. $\hat{V}_{-1,m}=\hat{U}_{2N_l+2,m}=\hat{P}_{2N_l+2,m}=0$. For consistency with the classical solutions presented above, we again adopt a Gaussian zonal forcing profile, whose Fourier transform is
\begin{equation}
	\hat{\mu}_0(k)=\sqrt{2\pi}\mathrm{e}^{-k^2/2}.
\end{equation}

 In practice, we supply a forcing whose vertical structure projects directly onto a single baroclinic mode corresponding to the replacement $F_m\to F_m\delta_{mq}$, where $q$ denotes the baroclinic mode being forced. The resulting finite linear system is straightforward to invert numerically for each zonal wavenumber. The corresponding physical-space fields are recovered using the inverse fast Fourier transform. 

\subsection{Results}

Steady-state solutions to (\ref{eq:firstbcmodesteady}) are shown alongside their classical uncoupled counterparts in fig.~\ref{fig:steadyplot}. Two distinct forcing profiles are chosen, with first and second baroclinic vertical structures in the top and bottom rows, respectively. The asymptotically re-scaled background filling fraction is chosen as $\overline{\mu}=1/2$, corresponding to an actual background filling fraction of approximately $3\%$. This is broadly consistent with small cloud-core area fractions typically found in cloud-resolving simulations \citep[see e.g.][]{siebesma1995,jakob2003}.

As might be expected, the response is dominated by the baroclinic mode in which the forcing is applied, meaning that the classical uncoupled solution captures the principal structure of the circulation remarkably well. The close agreement between the classical and coupled solutions provides strong support for the convective momentum transport parameterization developed here -- it reproduces essential leading-order features of the tropical circulation whilst providing a physically based description of the underlying convective process. The coupled solutions nevertheless exhibit systematic excitation of other baroclinic modes. The strength of this coupling is controlled by the background filling fraction together with the coupling matrix $\mathbf{K}$ whose form is determined from the small-scale convective structures. These additional mode contributions produce subtle modifications to the vertical structure of the circulation, providing a mechanism by which convection can redistribute momentum between vertical modes. Although the secondary responses have a modest amplitude compared with the directly forced mode, they consistently modify the phase of the vertical structure. 


Let us consider the meridionally integrated continuity equation
\begin{equation}\label{eq:meridionallyintegratedcont}
	\partial_X\overline{U}+\partial_z\overline{W}=0,
\end{equation}
where 
\begin{equation}
	\overline{U}(X,z)=\int_{-\infty}^\infty U(X,Y,z)\;\mathrm{d}Y,\qquad \overline{W}(X,z)=\int_{-\infty}^\infty W(X,Y,z)\;\mathrm{d}Y.
\end{equation}
Equation (\ref{eq:meridionallyintegratedcont}) permits the introduction of a Walker-circulation streamfunction $\overline{\Psi}$ defined by the relations $\overline{U}=-\overline{\Psi}_z$, $\overline{W}=\overline{\Psi}_X$. Using the vertical mode decomposition of $\overline{U}$, and requiring that the streamfunction vanish at the horizontal boundaries, we obtain
\begin{equation}\label{eq:walkersf}
	\overline{\Psi}(X,z)=-\sum_{m=1}^{N_m} \frac{\overline{U}_m(X)}{m}\sin(mz).
\end{equation}

\begin{figure}
	\centering
	\includegraphics[width=\textwidth]{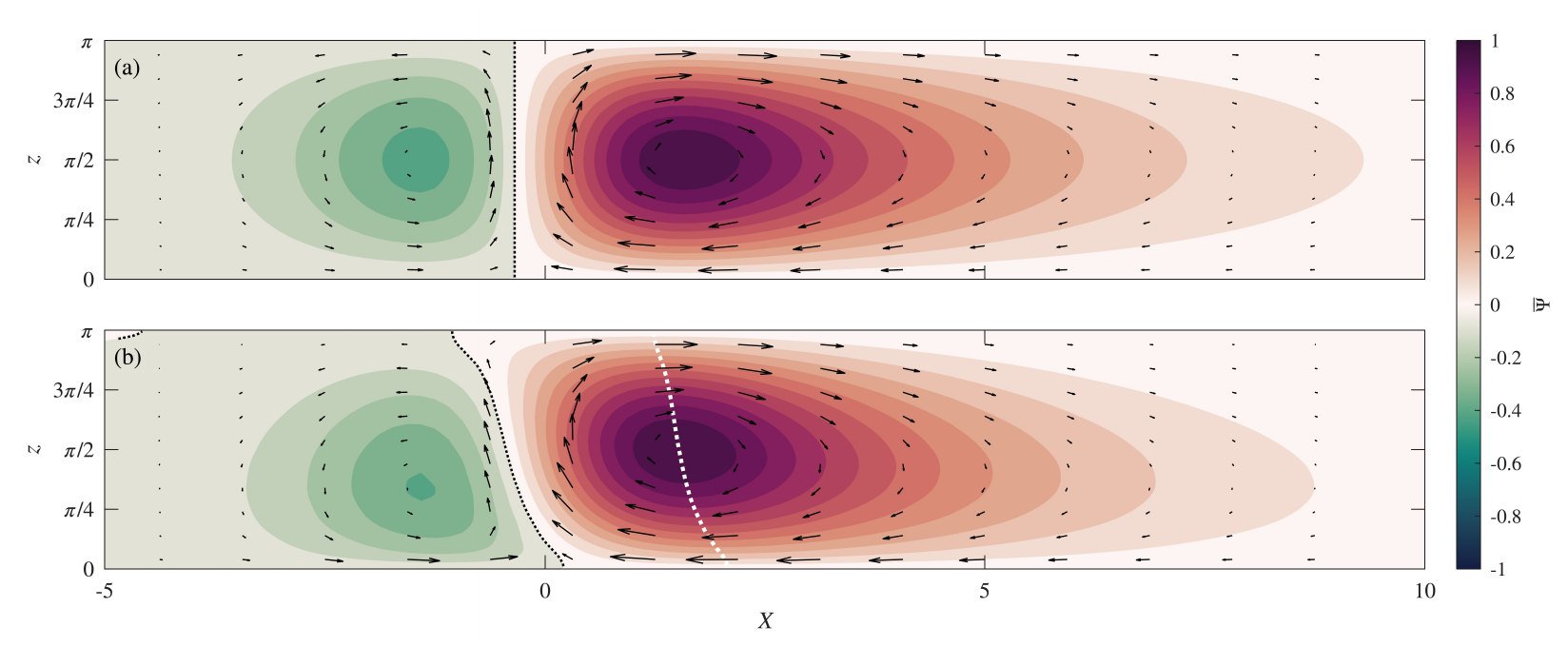}
	\caption{Walker-circulation streamfunction $\overline{\Psi}(X,z)$ for first baroclinic mode forcing with background filling fraction $\overline{\mu}=1/2$. Panel (a) shows the uncoupled solution and panel (b) the fully coupled solution. Arrows indicate the corresponding meridionally integrated circulation $(\overline{U},\overline{W})$. The black dotted line denotes the zero contour of $\overline{\Psi}$ whilst the white dotted line in panel (b) traces the horizontal location of the streamfunction maximum at each height. The streamfunction is normalized to unity across both panels. The numerical computations use $N_l=10$ meridional modes and $N_m=15$ baroclinic modes.}
	\label{fig:walkercirculation}
\end{figure}

Fig.~\ref{fig:walkercirculation} compares the Walker-circulation streamfunction for the classical uncoupled solution with that of the fully coupled system. Both solutions reproduce the main qualitative features of the Pacific Walker circulation, comprising a dominant basin-scale overturning cell located to the east of the convective anomaly together with a weaker secondary circulation of substantially shorter horizontal extent to the west. Despite their similar large-scale structure, the coupled and uncoupled solutions differ in the vertical organization of the primary overturning cell. In the classical solution (fig.~\ref{fig:walkercirculation}a), the center of the circulation remains vertically aligned throughout the troposphere. In contrast, the fully coupled solution (fig.~\ref{fig:walkercirculation}b) exhibits a systematic westward tilt with height. The tilt arises through the weak excitation of higher baroclinic modes by the convectively induced coupling operator. Although these higher modes have amplitudes that are small compared with the directly forced mode, they modify the phase of the overall vertical structure, producing a systematic westward deflection of the circulation center with increasing height. The white dotted curve denotes the zonal position of the streamfunction maximum at each height and provides a convenient measure of the circulation center. We see that the tilt is strongest in the lower troposphere, with dimensional slopes of approximately $100-175\mathrm{km}\;\mathrm{km}^{-1}$, decreasing to approximately $50-75\mathrm{km}\;\mathrm{km}^{-1}$ in the upper troposphere. These results demonstrate that convectively induced coupling between baroclinic modes can substantially modify the vertical structure of the Walker circulation whilst leaving its large-scale horizontal organization largely unchanged. The resulting westward tilt is qualitatively consistent with observations of the Pacific Walker circulation  \citep{hartmann1984,schumacher2004,iiponen2021}, suggesting that this coupling mechanism may contribute to its vertical organization.

\section{Tropical wave dynamics}\label{sec:waves}
To examine the effect of convectively induced coupling on the full equatorial wave spectrum, we now return to the coupled system (\ref{eq:coupledaveragedeqs}) in the absence of the equatorial long-wave approximation. Linearizing about a state of radiative equilibrium, neglecting perturbations about the background filling fraction $\overline{\mu}$, and seeking normal-mode solutions of the form $(U,V,P)=(\hat{U}(Y),\hat{V}(Y),\hat{P}(Y))\exp(\mathrm{i}kX-\mathrm{i}\omega t)$, we obtain
\begin{subequations}\label{eq:dispersionequations}
	\begin{align}
		-\mathrm{i}\omega \hat{U}_m-Y\hat{V}_m&=-\mathrm{i}k\hat{P}_m+\overline{\mu}\sum_{n=1}^{N_m} K_{mn}\hat{U}_n,\\
		-\mathrm{i}\omega \hat{V}_m+Y\hat{U}_m&=-\partial_Y\hat{P}_m+\overline{\mu}\sum_{n=1}^{N_m} K_{mn}\hat{V}_n,\\
		-\mathrm{i}\omega \hat{P}_m+\tfrac{1}{m^2}\left(\mathrm{i}k \hat{U}_m+\partial_Y\hat{V}_m\right)&=0,
	\end{align}
\end{subequations}
where, for simplicity, the buoyancy damping term is omitted. We compare the dispersion relations admitted by the coupled system (\ref{eq:dispersionequations}) with those of two corresponding uncoupled systems: first, the classical undamped equatorial waves obtained in the absence of small-scale convection ($\overline{\mu}=0$), and second, the diagonally damped approximation obtained by retaining only the diagonal components of $K_{mm}$. As in the steady-state analysis, the fully coupled system must be treated numerically, whereas the uncoupled systems admit analytical solutions.

\subsection{Uncoupled waves: analytical solutions}
Retaining only the diagonal components of $\mathbf{K}$, we define the 
effective momentum damping of the $m$th vertical mode by 
$-\overline{\mu}K_{mm}\equiv\kappa_m$. The governing equations for each 
vertical mode then decouple and reduce to
\begin{subequations}\label{eq:classicalwaves}
	\begin{align}
		(\kappa_m-\mathrm{i}\omega)\hat{U}_m-Y\hat{V}_m
		&=-\mathrm{i}k\hat{P}_m,\\
		(\kappa_m-\mathrm{i}\omega)\hat{V}_m+Y\hat{U}_m
		&=-\partial_Y\hat{P}_m,\\
		-\mathrm{i}\omega\hat{P}_m
		+\tfrac{1}{m^2}\left(\mathrm{i}k\hat{U}_m
		+\partial_Y\hat{V}_m\right)&=0.
	\end{align}
\end{subequations}
Eliminating $\hat{U}_m$ and $\hat{P}_m$, equations (\ref{eq:classicalwaves}) reduce to a single second-order equation for $\hat{V}_m$, given by 
\begin{equation}\label{eq:vhateq}
	\frac{d^2\hat{V}_m}{dY^2}+\left[m^2\omega(\omega+\mathrm{i}\kappa_m)-k^2-\frac{k}{\omega+\mathrm{i}\kappa_m}-\frac{m^2\omega}{(\omega+\mathrm{i}\kappa_m)}Y^2\right]\hat{V}_m=0.
\end{equation}
Equation (\ref{eq:vhateq}) admits non-trivial bounded solutions provided that the dispersion relation
\begin{equation}\label{eq:classicaldisprel}
	m^2\omega(\omega+\mathrm{i}\kappa_m)-k^2-\frac{k}{\omega+\mathrm{i}\kappa_m}=(2l+1)\sqrt{\frac{m^2\omega}{\omega+\mathrm{i}\kappa}},\qquad l=0,1,2,\dots
\end{equation}
is satisfied. The corresponding meridional eigenfunctions are given by the parabolic cylinder functions
\begin{equation}
	\hat{V}_m(Y)\propto \phi_l\left(\left[\frac{4m^2\omega}{\omega+\mathrm{i}\kappa_m}\right]^\frac{1}{4}Y\right),
\end{equation}
where the branch of the square root appearing in (\ref{eq:classicaldisprel}) is chosen to have positive real part, ensuring decay of $\hat{V}_m(Y)$ as $|Y|\to\infty$. The corresponding expressions for $\hat{U}_m(Y)$ and $\hat{P}_m(Y)$ may then be diagnosed straightforwardly from (\ref{eq:classicalwaves}).
The trivial solution $\hat{V}_m(Y)\equiv0$ of (\ref{eq:vhateq}) is also physically meaningful, corresponding to the Kelvin wave. Imposing meridional decay selects the dispersion relation 
\begin{equation}\label{eq:kelvinwavedisprel}
	\omega=-\frac{\mathrm{i}\kappa_m}{2}+\mathrm{sgn}(k)\sqrt{k^2/m^2-\kappa_m^2/4},
\end{equation}
with $k^2/m^2>\kappa^2/4$, and the corresponding eigenfunctions are
\begin{equation}
	\left(\hat{U}_m(Y),\hat{V}_m(Y),\hat{P}_m(Y)\right)\propto \left(1,0,k/m^2\omega\right)\exp(-Y^2m^2\omega/2k).
\end{equation}

Equations (\ref{eq:classicaldisprel}) and (\ref{eq:kelvinwavedisprel}) together define the dispersion relations for the complete spectrum of equatorially trapped uncoupled waves. In the absence of damping (i.e. when $\kappa=0$), they reduce to the classical freely propagating dispersion relations of \cite{matsuno:1966,gill:1982}. For $l=1,2,3,\dots$, the cubic equation (\ref{eq:classicaldisprel}) yields two counter-propagating inertio-gravity wave branches, and a single westward-propagating Rossby wave branch, whilst for $l=0$ the mixed Rossby--gravity wave is recovered. Equation (\ref{eq:kelvinwavedisprel}) describes the eastward-propagating Kelvin waves.

\begin{figure}
	\centering
	\includegraphics[width=\textwidth]{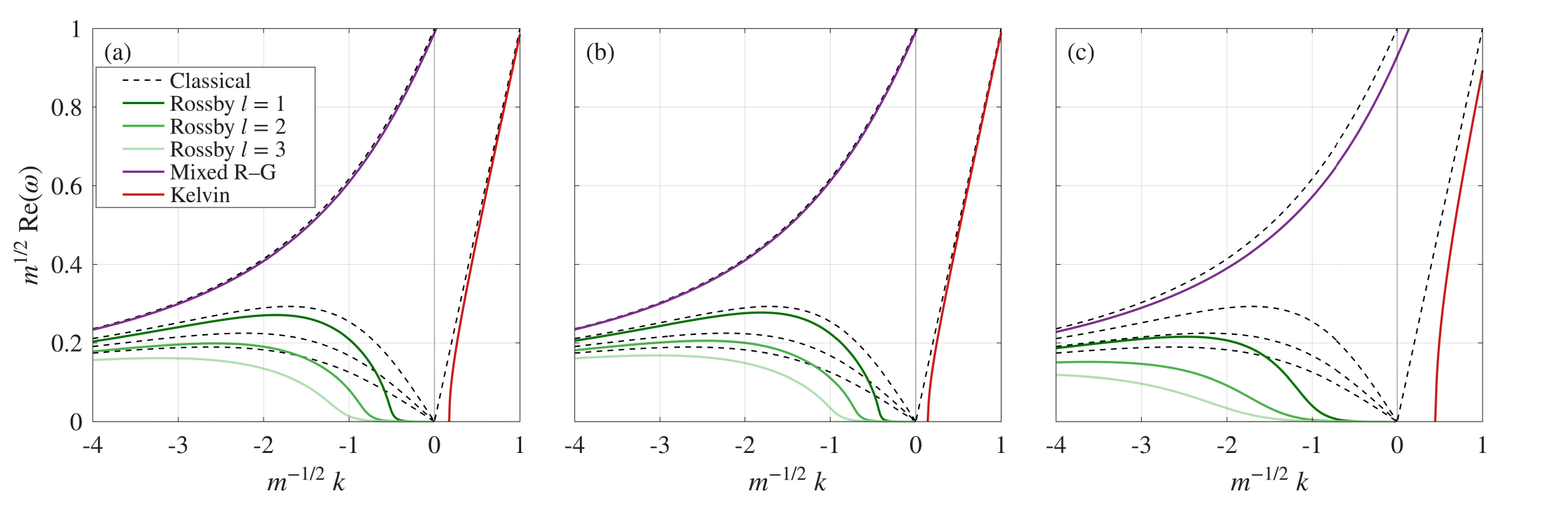}
	\caption{Dispersion relations for the uncoupled system (\ref{eq:classicalwaves}) showing low-frequency Rossby, mixed Rossby--gravity and Kelvin waves. Panels (a)--(c) show the first three vertical modes, with solid curves denoting the damped dispersion relations and dashed curves the corresponding classical undamped solutions. In each panel the frequency and zonal wavenumber are rescaled so that the classical dispersion relations are identical. Computations use $\overline{\mu}=1/2$ and $(K_{1\hspace{0.2mm}1},K_{2\hspace{0.2mm}2},K_{3\hspace{0.2mm}3})=(-0.691,-0.403,-1.039)$. }
	\label{fig:uncoupled_disprel}
\end{figure}

For the first three vertical modes examined here, damping has only a minor effect on the high-frequency wave branches, reflecting the fact that their intrinsic frequencies remain much larger than the corresponding damping rates. We therefore restrict our attention to the low-frequency spectrum. Fig.~\ref{fig:uncoupled_disprel} shows the dispersion relations for the uncoupled Rossby and low-frequency mixed Rossby--gravity waves computed from (\ref{eq:classicaldisprel}), together with the low-frequency Kelvin waves computed from (\ref{eq:kelvinwavedisprel}). In each case, damping systematically reduces the wave frequency relative to its classical undamped counterparts, with the lowest frequency branches most strongly affected. The magnitude of the frequency reduction varies between vertical modes in accordance with the effective damping rate $\kappa_m=-\overline{\mu}K_{mm}$. In particular, the second vertical mode, for which $\kappa_m$ is smallest, exhibits the weakest departure from the classical dispersion curves, whereas the third vertical mode is modified most strongly. An additional consequence of damping is the introduction of finite cut-off wavenumbers for the Rossby and Kelvin branches. Unlike the classical dispersion relations, which admit Rossby and Kelvin waves arbitrarily close to $k=0$, damping introduces a minimum wavenumber required for wave propagation, with the branches terminating as their frequencies vanish. 

\subsection{Coupled waves: numerical framework}\label{sec:coupledwaves}
To examine the dispersion relations for the fully coupled system
(\ref{eq:dispersionequations}), we once again decompose the meridional
structure into a basis of parabolic cylinder functions. Whereas in the
steady-state problem the equatorial symmetry of the prescribed convective
anomaly allowed us to retain only even modes for $\hat{U}_m$ and
$\hat{P}_m$ and odd modes for $\hat{V}_m$, such a restriction would discard
the antisymmetric part of the wave spectrum. We therefore employ the
staggered expansions
\begin{equation}\label{eq:waveparaboliccylinderfunctionexpansion}
	\left[\hat{U}_m,\hat{P}_m\right](k,Y)
	=\sum_{l=0}^{N_l}
	[\hat{U}_{l,m},\hat{P}_{l,m}](k)\phi_l(Y),
	\qquad
	\hat{V}_m(k,Y)
	=\sum_{l=0}^{N_l+1}\hat{V}_{l,m}(k)\phi_l(Y).
\end{equation}
Using the identities (\ref{eq:pcfidentities}) and projecting onto the
retained basis functions then gives the eigenvalue problem
\begin{subequations}\label{eq:coupledevproblem}
	\begin{align}
		\mathrm{i}\sqrt{\tfrac{l}{2}}\hat{V}_{l-1,m}
		+\mathrm{i}\sqrt{\tfrac{l+1}{2}}\hat{V}_{l+1,m}
		+k\hat{P}_{l,m}
		+\mathrm{i}\overline{\mu}\sum_{n=1}^{N_m}K_{mn}\hat{U}_{l,n}
		&=\omega\hat{U}_{l,m},
		\\
		-\mathrm{i}\sqrt{\tfrac{l}{2}}\hat{U}_{l-1,m}
		-\mathrm{i}\sqrt{\tfrac{l+1}{2}}\hat{U}_{l+1,m}
		+\mathrm{i}\sqrt{\tfrac{l}{2}}\hat{P}_{l-1,m}
		-\mathrm{i}\sqrt{\tfrac{l+1}{2}}\hat{P}_{l+1,m}
		+\mathrm{i}\overline{\mu}\sum_{n=1}^{N_m}K_{mn}\hat{V}_{l,n}
		&=\omega\hat{V}_{l,m},
		\\
		\tfrac{1}{m^2}\left(
		k\hat{U}_{l,m}
		+\mathrm{i}\sqrt{\tfrac{l}{2}}\hat{V}_{l-1,m}
		-\mathrm{i}\sqrt{\tfrac{l+1}{2}}\hat{V}_{l+1,m}
		\right)
		&=\omega\hat{P}_{l,m}.
	\end{align}
\end{subequations}
Equations (\ref{eq:coupledevproblem}a,c) are imposed for
$l=0,\dots,N_l$, whilst (\ref{eq:coupledevproblem}b) is imposed for
$l=0,\dots,N_l+1$. All coefficients whose indices lie outside the
corresponding retained ranges are taken to vanish.

\begin{figure}
	\centering
	\includegraphics[width=0.9\textwidth]{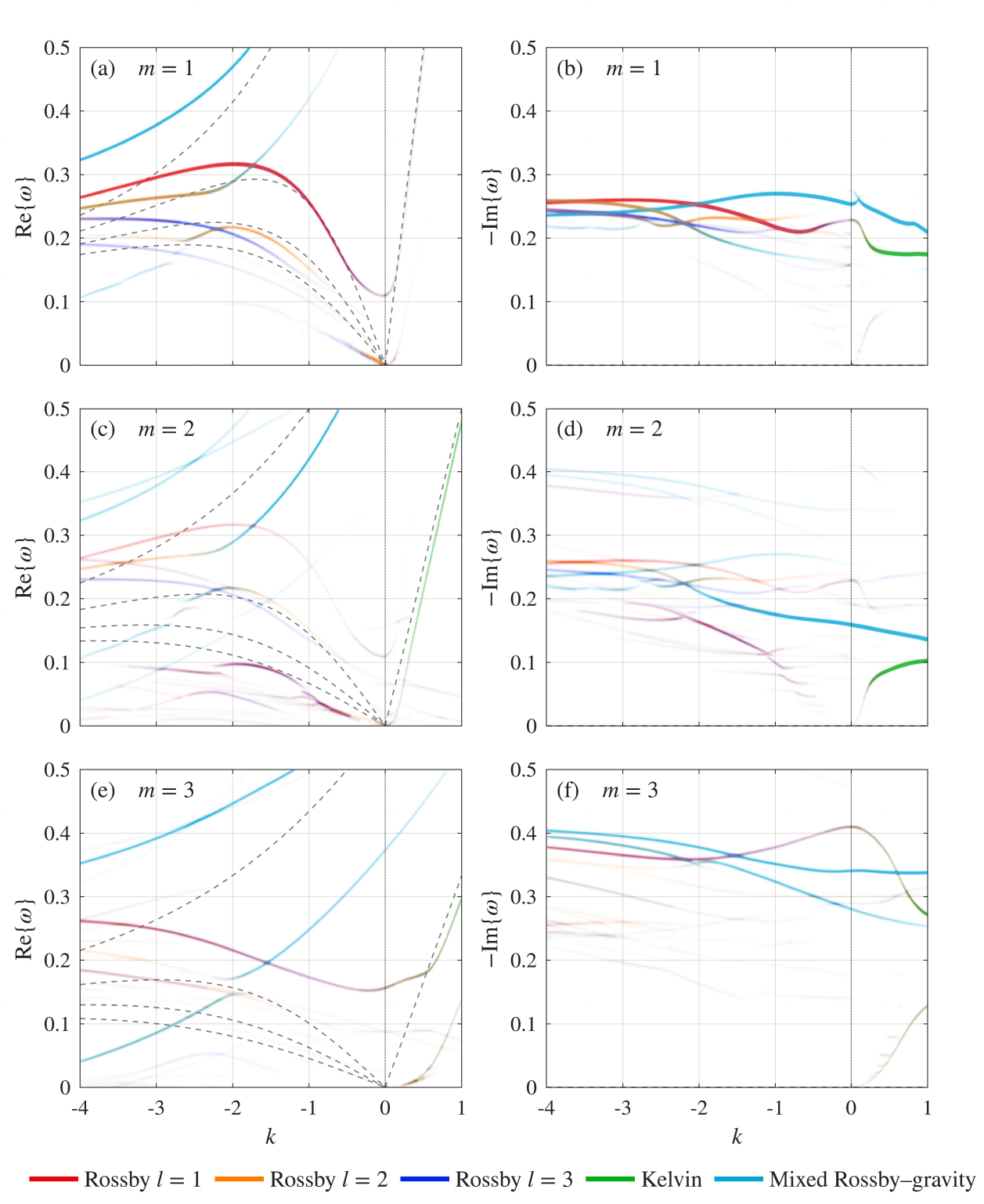}
	\caption{Reconstructed modal-content fields $C^{(a)}*{m,\mathrm{Re}}(k,\omega)$ and $C^{(a)}*{m,\gamma}(k,\gamma)$, where $\gamma=-\mathrm{Im}\{\omega\}$, constructed from the eigenmodes of the coupled system (\ref{eq:dispersionequations}). Panels (a,b), (c,d) and (e,f) show projections of the common coupled eigenspectrum onto uncoupled reference structures associated with the vertical modes ($m=1,2,3$), respectively. The left and right columns show the oscillation frequencies $\mathrm{Re}\{\omega\}$ and decay rates $-\mathrm{Im}\{\omega\}$. Modal content is coloured according to its projection onto the uncoupled Rossby $(l=1,2,3)$, Kelvin and mixed Rossby--gravity wave structures. Individual projections with content $\leq5\%$ are omitted before reconstruction. Opacity denotes the aggregate reconstructed content, while colour blending indicates simultaneous projection onto multiple reference structures. Grey dashed curves in the left column show the corresponding uncoupled dispersion relations; the uncoupled decay rates are zero. The same modal-content weights are used to construct the frequency and decay-rate representations of each eigenmode. A common opacity normalization is applied throughout the figure, defined by the largest of the $99.5$th-percentile aggregate-content values obtained from the three real-frequency rows. Computations use $\overline{\mu}=0.5$, with vertical and parabolic-cylinder truncations $N_m=15$ and $N_l=28$, respectively.
	}
	\label{fig:coupled_disprel}
\end{figure}

The eigenvectors of (\ref{eq:coupledevproblem}) corresponding to the eigenvalue $\omega$, represent oscillating wave structures whose frequency is given by $\mathrm{Re}(\omega)$ and whose decay rate is given by $-\mathrm{Im}(\omega)$. Owing to the coupling between vertical modes, the associated eigenvectors generally comprise mixtures of the classical equatorial wave structure. Consequently, the dispersion relations no longer organize into distinct branches with uniquely identifiable wave structures, as they do in the uncoupled system. This makes the spectrum inherently difficult to visualize. Fig.~\ref{fig:coupled_disprel}  presents a graphical representation of this spectrum in which each coupled eigenmode is classified according to its dominant uncoupled wave content.

To define the quantities represented in fig.~\ref{fig:coupled_disprel}, we introduce a reconstructed modal-content field based on projections of the coupled eigenvectors onto reference eigenvectors of the uncoupled system.
Let $\bs{q}_j(k)$ denote the $j$th eigenvector of (\ref{eq:coupledevproblem}) at wavenumber $k$, with associated complex eigenfrequency $\omega_j(k)$. We further let $\bs{\psi}_m^{(a)}(k)$ represent a reference eigenvector of the uncoupled system for vertical mode $m$ and wave type 
\begin{equation}
	\begin{aligned}
		a\in\{
		\mathrm{Rossby}\ (l=1),\;
		\mathrm{Rossby}&\ (l=2),\;
		\mathrm{Rossby}\ (l=3),\\
		&\mathrm{Kelvin},\;
		\mathrm{mixed\ Rossby\text{--}gravity}
		\}.
	\end{aligned}
\end{equation}
The uncoupled eigenvectors, obtained from (\ref{eq:classicalwaves}) in the undamped limit, are embedded in the $m$th vertical-mode block of the full coupled state space, with all other vertical-mode blocks set to zero.
Both sets of eigenvectors are normalized according to the weighted inner product $\langle \bs{\alpha},\bs{\beta}\rangle_\mathcal{W}=\bs{\alpha}^\dagger\mathcal{W}\bs{\beta}$, where the weight matrix $\mathcal{W}$ is given by
\begin{equation}
	\mathcal{W}=\begin{pmatrix}
		I_U &0 &0\\
		0&I_V& 0\\
		0&0&M_P^2
		\end{pmatrix}.
\end{equation}
Here $I_U$ and $I_V$ denote identity matrices on the full zonal and meridional velocity coefficient spaces, respectively, whilst \begin{equation}
	M_P^2
	=
	\operatorname{diag}(1^2,2^2,\ldots,N_m^2)\otimes I_P 
\end{equation}
supplies the vertical-mode weighting of the pressure coefficients, with $I_P$
denoting the identity on the parabolic-cylinder pressure-coefficient space. The normalized eigenvectors therefore satisfy $\langle\bs{q}_j,\bs{q}_j\rangle_\mathcal{W}=\langle \bs{\psi}_m^{(a)},\bs{\psi}_m^{(a)}\rangle_\mathcal{W}=1$.

For a coupled eigenvector $\bs{q}_j(k)$, its content associated with uncoupled wave type $a$ and vertical mode $m$ is defined as
\begin{equation}
	c_{j,m}^{(a)}(k)=\left|\left\langle\bs{\psi}^{(a)}_m(k),\bs{q}_j(k)\right\rangle_\mathcal{W}\right|^2.
\end{equation}
Since both vectors have unit weighted norm, we have $0\leq c^{(a)}_{j,m}(k)\leq 1$, where a value close to $1$ indicates that the coupled eigenvector closely resembles that particular uncoupled wave structure, whilst a value close to zero means that it contains very little of that structure. Each individual content with value $c_{j,m}^{(a)}\leq 0.05$ is discarded so that weak projections are removed. Importantly, these contents are not necessarily mutually exclusive fractions. The five reference vectors are not collectively orthogonalized, so a coupled eigenvector may have appreciable content in multiple reference structures. Their contents are therefore treated as individually labeled overlaps rather than as a partition of unity, and their sum is not assumed to equal 1. Schematically, the real-frequency content field is constructed as 
\begin{equation}
	C_{m,\mathrm{Re}}^{(a)}(k,\omega)=\sum_j c_{j,m}^{(a)}(k)G_{\sigma_\omega}(\omega-\mathrm{Re}\{\omega_j(k)\}),
\end{equation}
where $G_{\sigma_\omega}(\xi)=(2\pi\sigma_\omega^2)^{-1/2}\exp(-\xi^2/2\sigma_\omega^2)$ is a narrow Gaussian with standard deviation $\sigma_\omega=2\times10^{-3}$. The decay-rate field is similarly constructed as
\begin{equation}
	C^{(a)}_{m,\mathrm{Im}}(k,\gamma)=\sum_jc^{(a)}_{j,m}(k)G_{\sigma_\omega}(\gamma+\mathrm{Im}\{\omega_j(k)\}),
\end{equation}
where $\gamma=-\mathrm{Im}\{\omega\}$. The five reconstructed fields are displayed in fig.~\ref{fig:coupled_disprel} as a single composite field. Their relative magnitudes determine the local colour mixture, whilst the sum $C^\Sigma_{m,\mathrm{Re}}=\sum_aC^{(a)}_{m,\mathrm{Re}}(k,\omega)$ defines the aggregate reconstructed content used to determine opacity. To prevent exceptionally large values from dominating the opacity scale, $C^\Sigma_{m,\bullet}$ is normalized using its $99.5$th percentile for each vertical mode $m$. The same normalization factor is applied to the corresponding decay-rate field, so that equal aggregate content is represented by equal opacity in both columns. The parabolic cylinder and baroclinic mode truncations are set at $N_l=28$ and $N_m=15$, respectively, since the spectrum is well converged at these values (see appendix~\ref{app:convergence}).

In discussing the results shown in fig.~\ref{fig:coupled_disprel}, we use R1, R2, R3 to refer to content associated with the first three meridional Rossby modes, K the kelvin mode, and Y the mixed Rossby-gravity (Yanai) mode. The same coupled eigenspectrum underlies all three rows; the rows differ only in the vertical-mode reference structures onto which the associated eigenvectors are projected. In particular, the dominant branches carrying R1, R2, R3 and Y content in the $m=1$ row are the same as those appearing most prominently in the $m=2$ row. The associated eigenvectors therefore carry substantial contributions from both vertical modes and may not be assigned uniquely to either $m=1$ or $m=2$. Despite this vertical mixing, the dominant branches remain coherent and recognizable through their projections onto the classical wave structures. They are systematically displaced towards larger oscillation frequencies relative to the classical dispersion relations. This behavior is contrary to that observed in the diagonal damping case (see fig.~\ref{fig:uncoupled_disprel}), where frequencies are systematically shifted downwards, indicating that the off-diagonal vertical-mode coupling has an important dispersive effect in addition to producing decay.

The redistribution of wave character appears to be organized principally by equatorial parity. In the even-$l$ parity sector, a secondary branch carrying Y content hybridizes with the R2 principal branch. This behaviour is observed for all three vertical projections, although it is somewhat less prominent for $m=3$. Similarly, in the odd-$l$ parity sector, the R1 and R3 contents also become mixed. In the $m=1$ projection, this is seen most strongly in the principal R3 branch, whose magenta colouring arises through this mixing. In the $m=2$ projection, the R1-R3 hybridization is seen most prominently spread across several low-frequency magenta branches. In the $m=3$ projection, it is seen that almost all R1 and R3 content is concentrated into a single magenta branch, whilst the R2 content is fragmented across numerous weaker branches. Thus, neither the vertical-mode index $m$ nor the individual meridional index $l$ provides a global classification of the coupled eigenmodes. Equatorial parity, however remains an effective organizing principle.

The principal K content branch remains comparatively close to the uncoupled Kelvin dispersion relation over most of the positive wavenumber range. At small positive wavenumbers, however, eigenmodes carrying Kelvin content also acquire appreciable R1 and R3 content, producing low-frequency, long-wavelength Kelvin-Rossby hybrid structures. These structures appear in all three vertical-mode projections but have especially prominent $m=3$ content. Their preferential association with the odd-parity Rossby structures is again consistent with organization by equatorial symmetry. Under the dimensional scales employed here, these modes occupy planetary spatial scales, intraseasonal time scales and propagate slowly eastward. Their combined Kelvin-Rossby character is therefore suggestive of aspects of intraseasonal variability \citep{wang2018}. This resemblance should not be interpreted as identifying the modes with the Madden-Julian Oscillation, since the present linear model does not include interactive moisture and convective feedbacks required for a complete MJO mode \citep{sobel2013,jiang2020}. Nevertheless, it suggests that vertically non-local convective momentum coupling may contribute to the dynamical organization of such structures.

All eigenfrequencies in the figure have negative imaginary parts and are therefore damped. Because the rows show projections of the same eigenspectrum, the right-hand panels should not be interpreted as separate damping spectra for $m=1,2,3$. Rather they show the decay rates of those common eigenmodes that carry appreciable content in each vertical-mode subspace. The eigenmodes carrying strongest $m=1$ content have decay rates concentrated primarily within $0.15<\gamma<0.3$. Appreciable $m=2$ content is distributed among modes with decay rates spanning $0.05<\gamma<0.3$, while appreciable $m=3$ content occurs over the full range of $\gamma$ values shown. The increasing spread therefore indicates that the higher vertical-mode components are distributed among eigenmodes with increasingly heterogeneous decay rates, rather than that each vertical mode possesses a distinct or uniformly increasing damping coefficient.

Taken together, the results show that the off-diagonal coupling creates genuinely shared baroclinic eigenmodes. These modes span multiple vertical eigenmode subspaces, exchange classical wave character within the permitted parity classes, and experience branch-dependent frequency shifts and decay rates. The coupled spectrum therefore cannot be understood as a collection of independently damped classical equatorial waves associated with separate vertical modes. This behaviour contrasts with our earlier homogenization study of a midlatitude $\beta$-channel \citep{gol:23}, in which convection systematically reduced the frequencies of baroclinic Rossby waves. The difference reflects the distinct asymptotic scalings and convective circulations employed, as well as the inclusion of nonlocal coupling in both the momentum and buoyancy equations in the earlier model. Together, the two studies demonstrate that convection does not produce a universal frequency correction: both its magnitude and direction depend on the convective circulation and the form of coupling retained in the large-scale equations.

\section{Conclusions}\label{sec:conclusions}

In this paper, we have derived a multiple-scales theory describing the interaction between small-scale convection and the large-scale tropical circulation. Using asymptotic techniques based on the theory of homogenization, we obtained closed equations governing the large-scale equatorial flow in which the effects of convection enter through a non-local momentum diffusion term, derived directly from the underlying multiscale dynamics. This operator provides a first-principles representation of the convective momentum transport that has traditionally been represented phenomenologically in reduced theories of the tropical atmosphere. In the appropriate limiting case, the resulting closure reduces to the effective damping introduced by Gill, thereby placing the classical Matsuno--Gill framework on a systematic first-principles foundation.

The fully coupled system was shown to retain the characteristic Kelvin--Rossby structure of the classical Gill solution when subjected to a localized equatorially symmetric forcing. Unlike the uncoupled theory, however, convective momentum transport excites additional baroclinic modes beyond the directly forced mode. Although these secondary responses have comparatively small amplitudes, they exhibit systematic phase shifts with vertical wavenumber, collectively producing a westward tilt of the Walker circulation with height. This demonstrates that even relatively weak convectively induced coupling between vertical modes can produce systematic changes in the vertical structure of the large-scale tropical circulation which are qualitatively consistent with observations \citep{hartmann1984,schumacher2004,iiponen2021}. It was further shown that the convectively induced vertical mode coupling fundamentally alters the equatorial wave spectrum. Rather than decomposing into distinct Kelvin, Rossby, inertio--gravity and mixed Rossby--gravity wave families within individual baroclinic and meridional modes, the resulting eigenmodes comprise mixtures of the classical equatorial wave structures, with the mixing occurring predominantly between modes of the same equatorial symmetry. Consequently, the dispersion relations no longer separate into distinct wave branches, but instead transition continuously between the classical equatorial wave families. For the representative convective field considered here, the first-baroclinic wave branches are, for the most part, systematically shifted towards higher frequencies, whereas the second- and third-baroclinic branches exhibit more complex spectral behavior. In addition to this, we observe the presence of eastward-propagating Kelvin--Rossby hybrid structures reminiscent of those frequently invoked in theoretical descriptions of the Madden--Julian Oscillation \citep[see e.g.][]{bie:05,majda2009,majda2011}, suggesting that convectively-induced vertical-mode coupling may play a role in shaping intraseasonal tropical variability.

The present work suggests a different physical interpretation of the interaction between deep convection and the large-scale tropical circulation. Rather than acting simply as a local damping mechanism, unresolved convection gives rise to intrinsically non-local momentum transport that couples the baroclinic structure of the large-scale flow. More generally, this perspective is consistent with recent homogenization studies in other atmospheric settings, which likewise show that unresolved multiscale processes generate physically consistent non-local closures rather than ad hoc local parameterizations \citep{gol:23,gol02:25}. The phenomenological damping introduced by \cite{gill:1980} therefore emerges as the leading-order approximation of a more general non-local closure, placing the classical Matsuno--Gill framework within a broader hierarchy of systematically derived multiscale theories.

The theory developed in this paper nevertheless remains idealized. In particular, the convective field considered here represents a canonical model for deep convection, omitting the moist dynamics responsible for cloud formation and latent heat release. Consequently, the associated large-scale thermal forcing is prescribed rather than derived self-consistently from the underlying convective dynamics. Extending the present homogenization framework to incorporate more realistic moist convective processes therefore represents a promising avenue for future work, with the prospect of deriving both the thermal and momentum feedbacks directly from the underlying cloud dynamics.

\vspace{0.5cm}

\noindent\textbf{Acknowledgments.} The authors used ChatGPT (OpenAI, GPT-5.5) during the preparation of this manuscript between May and August 2026 to assist with revising and editing the text, and with debugging MATLAB code used in the numerical computations. The authors take full responsibility for all aspects of the manuscript, including the scientific content, mathematical derivations, numerical methods, interpretations, and conclusions.

\vspace{0.5cm}

\noindent\textbf{Funding.} Authors EJG and LD were funded through the LAUREATES 2026 summer research program at Hillsdale College. Additional support was provided by the Department of Mathematics at Hillsdale College. 
	
\vspace{0.5cm}
	
\noindent\textbf{Declaration of interests.} The authors report no conflicts of interest.

%
%
%
%
%
%

\small
\appendix

\section{Ertel's potential vorticity}\label{app:ErtelsPV}
For use in our multiple scales analysis, we derive Ertel's potential vorticity equation for the system (\ref{eq:startingeqns}). Let us define the planetary vorticity $\bs{\Omega}$ and total buoyancy $\Theta$ as
\refstepcounter{equation}
$$
\bs{\Omega}=\bs{\omega}+\beta y\zhat,\qquad \Theta=N^2z+b,
\eqno{(\theequation{\mathit{a},\mathit{b}})}
$$
where $\bs{\omega}=\nabla\times\bs{v}$ is the relative vorticity. Ertel's potential vorticity $q$ is defined by
\begin{equation}
	q=\bs{\Omega}\cdot\nabla\Theta,
\end{equation}
so that after taking the material derivative
\begin{equation}\label{eq:pvpv}
	\frac{Dq}{Dt}=\frac{D\bs{\Omega}}{Dt}\cdot\nabla\Theta+\bs{\Omega}\cdot\nabla\left(\frac{D\Theta}{Dt}\right)-\bs{\Omega}\cdot(\nabla\bs{v})^T\nabla\Theta.
\end{equation}
Taking the curl of the momentum equation (\ref{eq:startingeqsmom}), we find
\begin{equation}\label{eq:pvvorticity}
	\frac{D\bs{\Omega}}{Dt}=\bs{\Omega}\cdot\nabla\bs{v}+\nabla b\times\zhat-\lambda_m\bs{\omega},
\end{equation}
whilst the buoyancy equation becomes
\begin{equation}\label{eq:pvbuoyancy}
	\frac{D\Theta}{Dt}=S-\lambda_bb.
\end{equation} 
Substituting (\ref{eq:pvvorticity}) and (\ref{eq:pvbuoyancy}) into (\ref{eq:pvpv}), after simplification, yields
\begin{equation}\label{eq:ertelspv}
	q_t+\bs{v}\cdot\nabla q=\bs{\Omega}\cdot\nabla(S-\lambda_b b)-\lambda_m(\bs{\Omega}-\beta y\zhat)\cdot\nabla\Theta,
\end{equation}
which is Ertel's potential vorticity equation. In the absence of the buoyancy source, and momentum and buoyancy damping, equation (\ref{eq:ertelspv}) reduces to $Dq/Dt=0$, i.e. Ertel's potential vorticity is materially conserved.
\subsection{Asymptotics}
Non-dimensionalizing (\ref{eq:ertelspv}) using the vorticity scale $c/L_c$ and the potential vorticity scale $c^3/H^3$, substituting in the gradient operator (\ref{eq:gradientexpansion}), and adopting the distinguished limits for the damping coefficients and buoyancy source (\ref{eq:distinguishedlimits}), we find 
\begin{align}\label{eq:msPVeq}
	\eps\left(q_t+\bs{v}\cdot\onab q\right)+\bs{v}\cdot\nabla q=\eps^\frac{1}{2}\left(\eps\bs{\Omega}\cdot\onab +\bs{\Omega}\cdot\nabla\right)(S-\eps^\frac{1}{2}\gamma_bb)-\eps^\frac{1}{2}\gamma_m(\bs{\Omega}-\eps Y\zhat)\cdot\left(\eps\onab \Theta+\nabla \Theta\right),
\end{align}
where variables now correspond to their dimensionless counterparts, and
\refstepcounter{equation}
$$
q=\bs{\Omega}\cdot(\eps\onab\Theta+\nabla\Theta),\qquad\bs{\Omega}=\eps\onab\times\bs{v}+\nabla\times\bs{v}+\eps Y\zhat,\qquad \Theta=z+b.
\eqno{(\theequation{\mathit{a},\mathit{b},\mathit{c}})}
$$
Remaining consistent with the expansions (\ref{eq:perturbationexpansions}) and the axisymmetric leading-order flow profile (\ref{eq:leadingordervelocityfield}), we find the leading-order balance in equation (\ref{eq:msPVeq}) occurs at $O(\eps^{3/2})$, and is given by
\begin{equation}
	\bs{v}_0\cdot\nabla q_1=\bs{\Omega}_1\cdot\nabla S-\gamma_m(q_1-Y),
\end{equation}
where
\refstepcounter{equation}
$$
q_1=\zhat\cdot\nabla\times(\bs{v}_1+\bs{U})+Y,\qquad \bs{\omega}_1=\nabla\times(\bs{v}_1+\bs{U}), \qquad \bs{\Omega}_1=\nabla\times(\bs{v}_1+\bs{U})+Y\zhat.
\eqno{(\theequation{\mathit{a},\mathit{b},\mathit{c}})}
$$
With the $O(\eps)$ velocity field defined in terms of the stream function $\psi_1$ as in (\ref{eq:o1streamfunction}), Ertels potential vorticity and the absolute vorticity at leading order reduce to
\begin{subequations}
	\begin{align}
		q_1&=-\nabla^2_h \psi_1+Y,\\ \bs{\Omega}_1&=\left(\frac{\partial^2\psi_1}{\partial r\partial z}-\that\cdot\frac{\partial\bs{U}}{\partial z}\right)\rhat+\left(\frac{1}{r}\frac{\partial^2\psi_1}{\partial\theta\partial z}+\rhat\cdot\frac{\partial\bs{U}}{\partial z}\right)\that+\left(Y-\nabla^2_h\psi_1\right)\zhat,
	\end{align}
\end{subequations}
where $\nabla^2_h$ is the horizontal Laplacian.

\section{Planetary vorticity correction}\label{app:vorticitycorrection}
The convective response to planetary-vorticity forcing is governed by the system
\begin{subequations}\label{eq:planvortresponse}
	\begin{align}
		(u_0^r\partial_r+w_0\partial_z)\zeta_\Pi
		-\partial_r w_0\,\partial_{rz}^2\Pi
		-\partial_z w_0\,\zeta_\Pi
		+\gamma_m\zeta_\Pi
		&=\partial_z w_0,\\
		\zeta_\Pi
		&=-\partial_{rr}^2\Pi-\frac{1}{r}\partial_r\Pi,
	\end{align}
\end{subequations}
where the cloud-scale streamfunction associated with the induced recirculation is
$\psi_\Pi(r,z,Y)=Y\,\Pi(r,z)$. Importantly, this component of the streamfunction is locally axisymmetric on the small scales and therefore makes no contribution to the momentum-flux convergence term in (\ref{eq:averagedeqs}). Nevertheless, the solution to (\ref{eq:planvortresponse}) describes a distinct modification of the small-scale circulation arising from the planetary-vorticity effects and is therefore of physical interest in its own right. Equations (\ref{eq:planvortresponse}) are closely related to those solved by \cite{gol01:25} in their study of the influence of planetary rotation on cloud circulations.

\begin{figure}
	\centering
	\includegraphics[width=0.5\textwidth]{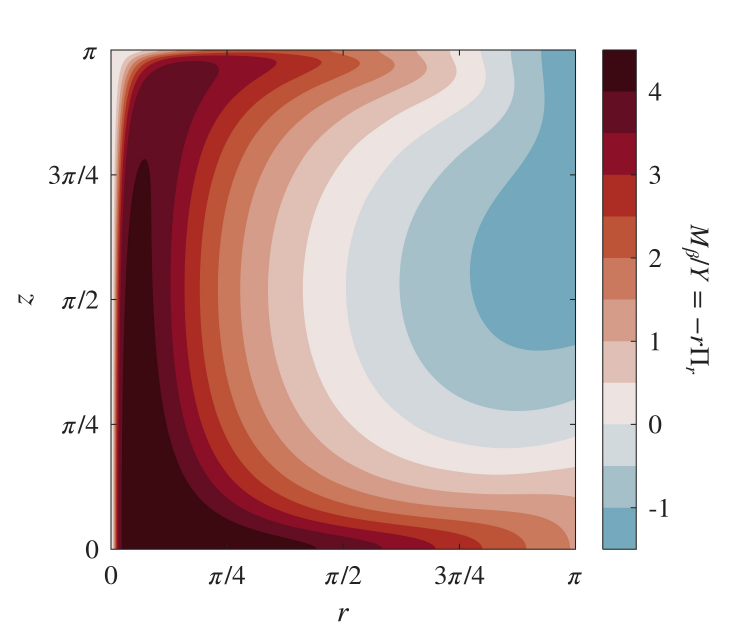}
	\caption{Specific relative angular momentum $M_\beta/Y=-r\Pi_r$ associated with the solution (\ref{eq:planvortresponse}).}
	\label{fig:pamresponse}
\end{figure}

Equations (\ref{eq:planvortresponse}) are solved analogously to (\ref{eq:cellprobem}) and (\ref{eq:modalcellproblem}), on the same computational domain and subject to the same boundary conditions on $z=0,\pi$ and decay conditions as $r\to\infty$. At the symmetry axis $r=0$, the regularity condition is replaced with the axisymmetric Neumann condition $\Pi_r(0,z)=0$. Since $\Pi(r,z)$ enters the flow only through the induced azimuthal velocity, we present the corresponding specific relative angular momentum,
\begin{equation}
	M_\beta/Y=-r\Pi_r,
\end{equation}
which provides a more direct measure of the rotational response. Fig.~\ref{fig:pamresponse} shows that planetary-vorticity forcing generates a localized axisymmetric rotational circulation confined to the convective region. The induced specific angular momentum is concentrated near the symmetry axis and decays rapidly with radial distance, indicating that the response remains localized to the cloud-scale circulation.

\section{Wave spectrum convergence} \label{app:convergence}
We assess the convergence of the wave spectrum shown in fig.~\ref{fig:coupled_disprel} with respect to the parabolic cylinder and baroclinic truncation numbers $N_l$ and $N_m$. To quantify convergence of the coupled spectrum, we compare the reconstructed spectral content distributions obtained using different truncations. Let $C_i(\omega,k)$ denote the spectral density associated with the $i$th diagnostic channel, constructed as in \S~\ref{sec:coupledwaves}, where the channels correspond to the R1, R2, R3, K and Y components within each of the first three vertical modes. For each wavenumber $k$, we define the integrated content 
\begin{equation}
	A_i(k)=\int C_i(\omega,k)\;\mathrm{d}\omega,
\end{equation}
and the normalized spectral distribution
\begin{equation}
	p_i(\omega,k)=\frac{C_i(\omega,k)}{A_i(k)}.
\end{equation}

\begin{figure}
	\centering
	\includegraphics[width=0.8\textwidth]{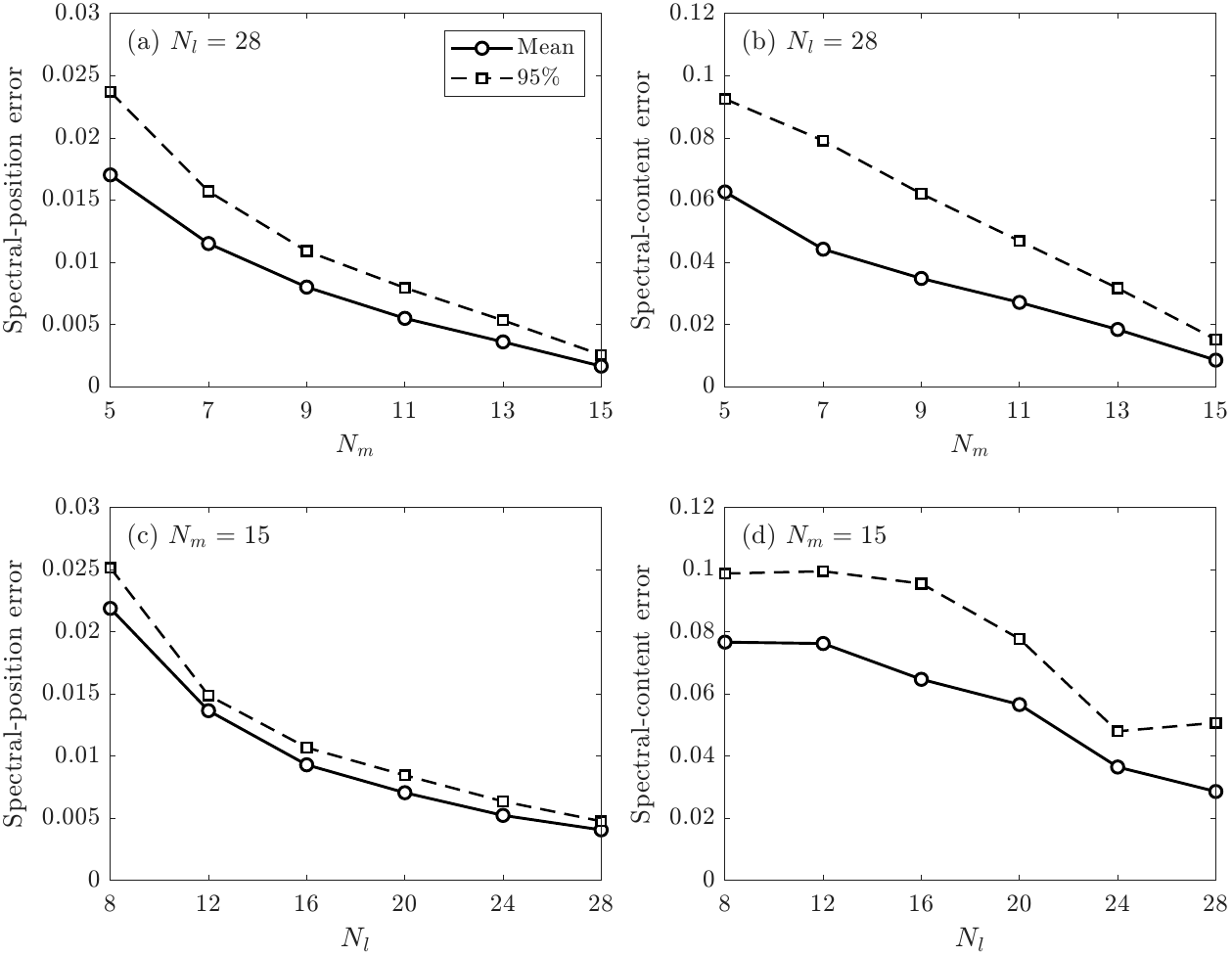}
	\caption{Convergence of the coupled wave spectrum with respect to the parabolic cylinder and baroclinic truncation errors $N_l$ and $N_m$. Panels (a,b) show convergence with increasing baroclinic truncation $N_m$ for fixed $N_l=28$, whilst panels (c,d) show convergence with increasing $N_l$ for fixed $N_m=15$. Panels (a,c) show the mean and 95th percentile of the spectral-position error computed from (\ref{eq:wasserstein}), whilst (b,d) show the mean and 95th percentile of the spectral-content error computed from (\ref{eq:contenterror}). The reference distributions use $N_m=17$ in (a,b) and $N_l=32$ in (c,d).}
	\label{fig:convergence}
\end{figure}

The first convergence metric measures the displacement of the spectral content using the one-dimensional Wasserstein distance \citep[see e.g.][]{villani2009},
\begin{equation}
	W_i(k)=\int \left| F_i(\omega,k)-F^\mathrm{ref}_i(\omega,k)\right|\;\mathrm{d}\omega,
\end{equation}
where $F_i(\omega,k)=\int^\omega p_i(\omega,k)\;\mathrm{d}\omega$ denotes the cumulative distribution associated with $p_i$, and the superscript ``ref'' denotes the highest-resolution calculation. The Wasserstein distance therefore measures the average displacement of the normalized spectral content in frequency space. A single measure of the spectral-position error is then obtained by averaging over all channels using the reference spectral content as weights,
\begin{equation}\label{eq:wasserstein}
	W(k)=\frac{\sum_i A_i^\mathrm{ref}(k)W_i(k)}{\sum_iA_i^\mathrm{ref}(k)}.
\end{equation}

The second metric measures the convergence of the integrated contents of the diagnostic channels,
\begin{equation}\label{eq:contenterror}
	E(k)=\frac{\sum_iA_i^\mathrm{ref}(k)\left|A_i(k)-A_i^\mathrm{ref}(k)\right|/A_i^\mathrm{ref}(k)}{\sum_iA_i^\mathrm{ref}(k)},
\end{equation}
which represents a weighted mean relative error in the integrated R1, R2, R3, K and Y wave contents.

For each truncation, we report both the mean and the 95th percentile of $W(k)$ and $E(k)$ over the full wavenumber range. Fig.~\ref{fig:convergence} demonstrates convergence of the coupled wave spectrum with increasing baroclinic truncation $N_m$ in panels (a) and (b), for fixed $N_l=28$, and with increasing parabolic cylinder truncation $N_l$ in panels (c) and (d), for fixed $N_m=15$. In both cases, the spectral-position and spectral-content error are seen to decrease monotonically as the truncation is increased. At the resolutions adopted in \S~\ref{sec:coupledwaves}, $N_m=15$ and $N_l=28$, both diagnostics are substantially reduced relative to coarser truncations, indicating the wave spectrum has converged with respect to both vertical and meridional discretization.

\newcommand{\noopsort}[1]{}


\begin{thebibliography}{52}
	\expandafter\ifx\csname natexlab\endcsname\relax\def\natexlab#1{#1}\fi
	\def\au#1{#1} \def\ed#1{#1} \def\yr#1{#1}\def\at#1{#1}\def\jt#1{\textit{#1}}
	\def\bt#1{#1}\def\bvol#1{\textbf{#1}} \def\vol#1{#1} \def\pg#1{#1}
	\def\publ#1{#1}\def\arxiv#1{#1}\def\org#1{#1}\def\st#1{\textit{#1}}
	
	\bibitem[Abramowitz \& Stegun(1964)]{abramowitz1964handbook}
	{\sc \au{Abramowitz, Milton} \& \au{Stegun, Irene~A.}}, ed. \yr{1964} {\em
		Handbook of Mathematical Functions with Formulas, Graphs, and Mathematical
		Tables\/},  \st{National Bureau of Standards Applied Mathematics Series},
	\vol{vol.~55}.  \publ{Washington, D.C.: U.S. Government Printing Office},
	reprinted by Dover Publications, New York, 1965.
	
	\bibitem[Allaire(2012)]{allaire2012}
	{\sc \au{Allaire, Gr{\'e}goire}} \yr{2012} A brief introduction to
	homogenization and miscellaneous applications.  \bt{In {\em ESAIM:
			Proceedings\/}}, ,  \vol{vol.~37},  \pg{pp. 1--49}. EDP Sciences.
	
	\bibitem[Arakawa \& Schubert(1974)]{arakawa:1974}
	{\sc \au{Arakawa, Akio} \& \au{Schubert, Wayne~Howard}} \yr{1974}
	\at{Interaction of a cumulus cloud ensemble with the large-scale environment,
		part i}.  \jt{Journal of Atmospheric Sciences}  \bvol{31}~(3),  \pg{674 --
		701}.
	
	\bibitem[Betts(1986)]{betts1986}
	{\sc \au{Betts, A.~K.}} \yr{1986}  \at{A new convective adjustment scheme. part
		i: Observational and theoretical basis}.  \jt{Quarterly Journal of the Royal
		Meteorological Society}  \bvol{112}~(473),  \pg{677--691}.
	
	\bibitem[Biello \& Majda(2010)]{bie:10}
	{\sc \au{Biello, Joseph} \& \au{Majda, Andrew}} \yr{2010}  \at{Intraseasonal
		multi-scale moist dynamics of the tropical atmosphere}.  \jt{Communications
		in Mathematical Sciences}  \bvol{8}.
	
	\bibitem[Biello \& Majda(2005)]{bie:05}
	{\sc \au{Biello, Joseph~A.} \& \au{Majda, Andrew~J.}} \yr{2005}  \at{A new
		multiscale model for the Madden--Julian Oscillation}.  \jt{Journal of the
		Atmospheric Sciences}  \bvol{62}~(6),  \pg{1694 -- 1721}.
	
	\bibitem[Biello \& Majda(2013)]{bie:13}
	{\sc \au{Biello, Joseph~A.} \& \au{Majda, Andrew~J.}} \yr{2013}  \at{A
		multiscale model for the modulation and rectification of the ITCZ}.
	\jt{Journal of the Atmospheric Sciences}  \bvol{70}~(4),  \pg{1053 -- 1070}.
	
	\bibitem[Boyd(1982)]{boyd1982}
	{\sc \au{Boyd, John~P}} \yr{1982}  \at{The optimization of convergence for
		chebyshev polynomial methods in an unbounded domain}.  \jt{Journal of
		Computational Physics}  \bvol{45}~(1),  \pg{43--79}.
	
	\bibitem[Boyd(2001)]{boy:01}
	{\sc \au{Boyd, J.~P.}} \yr{2001} {\em Chebyshev and {F}ourier {S}pectral
		{M}ethods\/}.  \publ{New York}.
	
	\bibitem[Cane {\em et~al.\/}(1986)Cane, Zebiak \& Dolan]{cane1986}
	{\sc \au{Cane, Mark~A.}, \au{Zebiak, Stephen~E.} \& \au{Dolan, Sean~C.}}
	\yr{1986}  \at{Experimental forecasts of {El Ni{\~n}o}}.  \jt{Nature}
	\bvol{321}~(6073),  \pg{827--832}.
	
	\bibitem[Emanuel(1986)]{emanuel1986}
	{\sc \au{Emanuel, Kerry~A.}} \yr{1986}  \at{An air-sea interaction theory for
		tropical cyclones. part i: Steady-state maintenance}.  \jt{Journal of
		Atmospheric Sciences}  \bvol{43}~(6),  \pg{585 -- 605}.
	
	\bibitem[Emanuel(1991)]{emanuel1991}
	{\sc \au{Emanuel, Kerry~A.}} \yr{1991}  \at{A scheme for representing cumulus
		convection in large-scale models}.  \jt{Journal of Atmospheric Sciences}
	\bvol{48}~(21),  \pg{2313 -- 2329}.
	
	\bibitem[Gill(1980)]{gill:1980}
	{\sc \au{Gill, A.~E.}} \yr{1980}  \at{Some simple solutions for heat-induced
		tropical circulation}.  \jt{Quarterly Journal of the Royal Meteorological
		Society}  \bvol{106}~(449),  \pg{447--462}.
	
	\bibitem[Gill(1982)]{gill:1982}
	{\sc \au{Gill, Adrian~E.}} \yr{1982} {\em Atmosphere--Ocean Dynamics\/}.
	\publ{New York: Academic Press}.
	
	\bibitem[Goldsmith {\em et~al.\/}(2025{\natexlab{{\em a\/}}})Goldsmith, Biello
	\& Igel]{gol01:25}
	{\sc \au{Goldsmith, Edward~J.}, \au{Biello, Joseph~A.} \& \au{Igel,
			Matthew~R.}} \yr{2025{\natexlab{{\em a\/}}}}  \at{Convective circulations and
		the Coriolis force: A mechanism for upscale momentum fluxes in the tropics}.
	\jt{Journal of the Atmospheric Sciences}  \bvol{82}~(5),  \pg{849 -- 867}.
	
	\bibitem[Goldsmith {\em et~al.\/}(2025{\natexlab{{\em b\/}}})Goldsmith, Biello
	\& Igel]{gol02:25}
	{\sc \au{Goldsmith, Edward~J.}, \au{Biello, Joseph~A.} \& \au{Igel,
			Matthew~R.}} \yr{2025{\natexlab{{\em b\/}}}}  \at{The nontraditional Coriolis
		force drives zonal shear in the tropical atmosphere: A systematic
		multiple-scales exposition}.  \jt{Journal of the Atmospheric Sciences}
	\bvol{82}~(10),  \pg{2237 -- 2253}.
	
	\bibitem[Goldsmith \& Esler(2023)]{gol:23}
	{\sc \au{Goldsmith, Edward~J.} \& \au{Esler, James~G.}} \yr{2023}  \at{Wave
		propagation through a stationary field of clouds: A homogenisation approach}.
	\jt{Quarterly Journal of the Royal Meteorological Society}
	\bvol{149}~(757),  \pg{3455--3476}.
	
	\bibitem[Hartmann {\em et~al.\/}(1984)Hartmann, Hendon \& Houze]{hartmann1984}
	{\sc \au{Hartmann, Dennis~L.}, \au{Hendon, Harry~H.} \& \au{Houze, Robert~A.}}
	\yr{1984}  \at{Some implications of the mesoscale circulations in tropical
		cloud clusters for large-scale dynamics and climate}.  \jt{Journal of
		Atmospheric Sciences}  \bvol{41}~(1),  \pg{113 -- 121}.
	
	\bibitem[Hoskins \& Karoly(1981)]{hoskins1981}
	{\sc \au{Hoskins, Brian~J.} \& \au{Karoly, David~J.}} \yr{1981}  \at{The steady
		linear response of a spherical atmosphere to thermal and orographic forcing}.
	\jt{Journal of Atmospheric Sciences}  \bvol{38}~(6),  \pg{1179 -- 1196}.
	
	\bibitem[Houze~Jr.(1989)]{houze1989}
	{\sc \au{Houze~Jr., Robert~A.}} \yr{1989}  \at{Observed structure of mesoscale
		convective systems and implications for large-scale heating}.  \jt{Quarterly
		Journal of the Royal Meteorological Society}  \bvol{115}~(487),
	\pg{425--461}.
	
	\bibitem[Houze~Jr.(2004)]{hou:04}
	{\sc \au{Houze~Jr., Robert~A.}} \yr{2004}  \at{Mesoscale convective systems}.
	\jt{Reviews of Geophysics}  \bvol{42}~(4).
	
	\bibitem[Iipponen \& Donner(2021)]{iiponen2021}
	{\sc \au{Iipponen, Juho} \& \au{Donner, Leo}} \yr{2021}  \at{Simple analytic
		solutions for a convectively driven walker circulation and their relevance to
		observations}.  \jt{Journal of the Atmospheric Sciences}  \bvol{78}~(1),
	\pg{299 -- 311}.
	
	\bibitem[Jakob \& Siebesma(2003)]{jakob2003}
	{\sc \au{Jakob, Christian} \& \au{Siebesma, A.~Pier}} \yr{2003}  \at{A new
		subcloud model for mass-flux convection schemes: Influence on triggering,
		updraft properties, and model climate}.  \jt{Monthly Weather Review}
	\bvol{131}~(11),  \pg{2765 -- 2778}.
	
	\bibitem[Jiang {\em et~al.\/}(2020)Jiang, Adames, Kim, Maloney, Lin, Kim,
	Zhang, DeMott \& Klingaman]{jiang2020}
	{\sc \au{Jiang, Xianan}, \au{Adames, Ángel~F.}, \au{Kim, Daehyun},
		\au{Maloney, Eric~D.}, \au{Lin, Hai}, \au{Kim, Hyemi}, \au{Zhang, Chidong},
		\au{DeMott, Charlotte~A.} \& \au{Klingaman, Nicholas~P.}} \yr{2020}
	\at{Fifty years of research on the Madden--Julian Oscillation: Recent
		progress, challenges, and perspectives}.  \jt{Journal of Geophysical
		Research: Atmospheres}  \bvol{125}~(17),  \pg{e2019JD030911}, e2019JD030911
	2019JD030911.
	
	\bibitem[Jung(2009)]{jun:09}
	{\sc \au{Jung, Jae-Hun}} \yr{2009}  \at{A note on the spectral collocation
		approximation of some differential equations with singular source terms}.
	\jt{Journal of Scientific Computing}  \bvol{39},  \pg{49--66}.
	
	\bibitem[Kucharski {\em et~al.\/}(2009)Kucharski, Bracco, Yoo, Tompkins,
	Feudale, Ruti \& Dell'Aquila]{kucharski2009}
	{\sc \au{Kucharski, F.}, \au{Bracco, A.}, \au{Yoo, J.~H.}, \au{Tompkins,
			A.~M.}, \au{Feudale, L.}, \au{Ruti, P.} \& \au{Dell'Aquila, A.}} \yr{2009}
	\at{A Gill–matsuno-type mechanism explains the tropical Atlantic influence
		on African and Indian monsoon rainfall}.  \jt{Quarterly Journal of the Royal
		Meteorological Society}  \bvol{135}~(640),  \pg{569--579}.
	
	\bibitem[Maher {\em et~al.\/}(2019)Maher, Gerber, Medeiros, Merlis, Sherwood,
	Sheshadri, Sobel, Vallis, Voigt \& Zurita-Gotor]{maher2019}
	{\sc \au{Maher, Penelope}, \au{Gerber, Edwin~P.}, \au{Medeiros, Brian},
		\au{Merlis, Timothy~M.}, \au{Sherwood, Steven}, \au{Sheshadri, Aditi},
		\au{Sobel, Adam~H.}, \au{Vallis, Geoffrey~K.}, \au{Voigt, Aiko} \&
		\au{Zurita-Gotor, Pablo}} \yr{2019}  \at{Model hierarchies for understanding
		atmospheric circulation}.  \jt{Reviews of Geophysics}  \bvol{57}~(2),
	\pg{250--280}.
	
	\bibitem[Majda(2003)]{majda2003}
	{\sc \au{Majda, Andrew~J.}} \yr{2003} {\em Introduction to PDEs and Waves for
		the Atmosphere and Ocean\/},  \st{Courant Lecture Notes in Mathematics},
	\vol{vol.~9}.  \publ{Providence, RI: American Mathematical Society}.
	
	\bibitem[Majda \& Klein(2003)]{maj:03}
	{\sc \au{Majda, Andrew~J.} \& \au{Klein, Rupert}} \yr{2003}  \at{Systematic
		multiscale models for the tropics}.  \jt{Journal of the Atmospheric Sciences}
	\bvol{60}~(2),  \pg{393 -- 408}.
	
	\bibitem[Majda \& Stechmann(2009)]{majda2009}
	{\sc \au{Majda, Andrew~J.} \& \au{Stechmann, Samuel~N.}} \yr{2009}  \at{The
		skeleton of tropical intraseasonal oscillations}.  \jt{Proceedings of the
		National Academy of Sciences}  \bvol{106}~(21),  \pg{8417--8422}.
	
	\bibitem[Majda \& Stechmann(2011)]{majda2011}
	{\sc \au{Majda, Andrew~J.} \& \au{Stechmann, Samuel~N.}} \yr{2011}
	\at{Nonlinear dynamics and regional variations in the MJO skeleton}.
	\jt{Journal of the Atmospheric Sciences}  \bvol{68}~(12),  \pg{3053 -- 3071}.
	
	\bibitem[Marsico {\em et~al.\/}(2023)Marsico, Biello \& Igel]{mar01:23}
	{\sc \au{Marsico, David~H.}, \au{Biello, Joseph~A.} \& \au{Igel, Matthew~R.}}
	\yr{2023}  \at{Balanced convective circulations in a stratified atmosphere.
		part i: A framework for assessing radiation, the Coriolis force, and drag}.
	\jt{Journal of the Atmospheric Sciences}  \bvol{80}~(12),  \pg{2915 -- 2924}.
	
	\bibitem[Matsuno(1966)]{matsuno:1966}
	{\sc \au{Matsuno, Taroh}} \yr{1966}  \at{Quasi-geostrophic motions in the
		equatorial area}.  \jt{Journal of the Meteorological Society of Japan}  \bvol{44}~(1),  \pg{25--43}.
	
	\bibitem[Neelin(1989)]{neelin:1989}
	{\sc \au{Neelin, J.~David}} \yr{1989}  \at{On the interpretation of the Gill
		model}.  \jt{Journal of Atmospheric Sciences}  \bvol{46}~(15),  \pg{2466 --
		2468}.
	
	\bibitem[Neelin(1991)]{neelin1991}
	{\sc \au{Neelin, J.~David}} \yr{1991}  \at{The slow sea surface temperature
		mode and the fast-wave limit: Analytic theory for tropical interannual
		oscillations and experiments in a hybrid coupled model}.  \jt{Journal of
		Atmospheric Sciences}  \bvol{48}~(4),  \pg{584 -- 606}.
	
	\bibitem[Neelin \& Zeng(2000)]{neelin2000}
	{\sc \au{Neelin, J.~David} \& \au{Zeng, Ning}} \yr{2000}  \at{A
		quasi-equilibrium tropical circulation model—formulation}.  \jt{Journal of
		the Atmospheric Sciences}  \bvol{57}~(11),  \pg{1741 -- 1766}.
	
	\bibitem[Pavliotis \& Stuart(2008)]{pavliotis:2008}
	{\sc \au{Pavliotis, Grigorios~A.} \& \au{Stuart, Andrew~M.}} \yr{2008} {\em
		Multiscale Methods: Averaging and Homogenization\/},  \st{Texts in Applied
		Mathematics},  \vol{vol.~53}.  \publ{New York: Springer}.
	
	\bibitem[Raymond \& Zeng(2005)]{raymond2005}
	{\sc \au{Raymond, David~J.} \& \au{Zeng, Xiping}} \yr{2005}  \at{Modelling
		tropical atmospheric convection in the context of the weak temperature
		gradient approximation}.  \jt{Quarterly Journal of the Royal Meteorological
		Society}  \bvol{131}~(608),  \pg{1301--1320}.
	
	\bibitem[Rodwell \& Hoskins(1996)]{rodwell1996}
	{\sc \au{Rodwell, Mark} \& \au{Hoskins, Brian}} \yr{1996}  \at{Monsoons and the
		dynamics of deserts}.  \jt{Quarterly Journal of the Royal Meteorological
		Society}  \bvol{122},  \pg{1385 -- 1404}.
	
	\bibitem[Schumacher {\em et~al.\/}(2004)Schumacher, Houze \&
	Kraucunas]{schumacher2004}
	{\sc \au{Schumacher, Courtney}, \au{Houze, Robert~A.} \& \au{Kraucunas, Ian}}
	\yr{2004}  \at{The tropical dynamical response to latent heating estimates
		derived from the trmm precipitation radar}.  \jt{Journal of the Atmospheric
		Sciences}  \bvol{61}~(12),  \pg{1341 -- 1358}.
	
	\bibitem[Siebesma \& Cuijpers(1995)]{siebesma1995}
	{\sc \au{Siebesma, A.~P.} \& \au{Cuijpers, J. W.~M.}} \yr{1995}  \at{Evaluation
		of parametric assumptions for shallow cumulus convection}.  \jt{Journal of
		Atmospheric Sciences}  \bvol{52}~(6),  \pg{650 -- 666}.
	
	\bibitem[Sobel \& Maloney(2013)]{sobel2013}
	{\sc \au{Sobel, Adam} \& \au{Maloney, Eric}} \yr{2013}  \at{Moisture modes and
		the eastward propagation of the MJO}.  \jt{Journal of the Atmospheric
		Sciences}  \bvol{70}~(1),  \pg{187 -- 192}.
	
	\bibitem[Sobel \& Bretherton(2000)]{sobel2000}
	{\sc \au{Sobel, Adam~H.} \& \au{Bretherton, Christopher~S.}} \yr{2000}
	\at{Modeling tropical precipitation in a single column}.  \jt{Journal of
		Climate}  \bvol{13}~(24),  \pg{4378 -- 4392}.
	
	\bibitem[Tiedtke(1989)]{tiedtke1989}
	{\sc \au{Tiedtke, M.}} \yr{1989}  \at{A comprehensive mass flux scheme for
		cumulus parameterization in large-scale models}.  \jt{Monthly Weather Review}
	\bvol{117}~(8),  \pg{1779 -- 1800}.
	
	\bibitem[Trefethen(2000)]{trefethenbook}
	{\sc \au{Trefethen, L.~N.}} \yr{2000} {\em {Spectral Methods in MATLAB}\/}.
	\publ{Philadelphia: SIAM}.
	
	\bibitem[Vallis(2017)]{vallisbook}
	{\sc \au{Vallis, Geoffrey~K.}} \yr{2017} {\em Atmospheric and Oceanic Fluid
		Dynamics: Fundamentals and Large-Scale Circulation\/}, 2nd edn.
	\publ{Cambridge University Press}.
	
	\bibitem[Villani(2009)]{villani2009}
	{\sc \au{Villani, C{\'e}dric}} \yr{2009} {\em Optimal Transport: Old and
		New\/},  \st{Grundlehren der mathematischen Wissenschaften},  \vol{vol. 338}.
	\publ{Springer}.
	
	\bibitem[Wang {\em et~al.\/}(2018)Wang, Li \& Nasuno]{wang2018}
	{\sc \au{Wang, Lu}, \au{Li, Tim} \& \au{Nasuno, Tomoe}} \yr{2018}  \at{Impact
		of Rossby and Kelvin wave components on MJO eastward propagation}.
	\jt{Journal of Climate}  \bvol{31}~(17),  \pg{6913 -- 6931}.
	
	\bibitem[Webster(1972)]{webster1972}
	{\sc \au{Webster, Peter~J.}} \yr{1972}  \at{Response of the tropical atmosphere
		to local, steady forcing}.  \jt{Monthly Weather Review}  \bvol{100}~(7),
	\pg{518 -- 541}.
	
	\bibitem[Whitaker(2013)]{whitaker:2013}
	{\sc \au{Whitaker, S.}} \yr{2013} {\em The Method of Volume Averaging\/}.
	\publ{Springer Netherlands}.
	
	\bibitem[Zhang(2005)]{zhang2005}
	{\sc \au{Zhang, Chidong}} \yr{2005}  \at{Madden-julian oscillation}.
	\jt{Reviews of Geophysics}  \bvol{43}~(2).
	
	\bibitem[Zhang \& Krishnamurti(1996)]{zhang1996}
	{\sc \au{Zhang, Z.} \& \au{Krishnamurti, T.~N.}} \yr{1996}  \at{A
		generalization of Gill's heat-induced tropical circulation}.  \jt{Journal of
		Atmospheric Sciences}  \bvol{53}~(7),  \pg{1045 -- 1052}.
	
\end{thebibliography}
\end{document}